\documentclass[%
 reprint,
 aps,prb,
 twocolumn,
 showpacs,superscriptaddress]
 {revtex4-1}

\usepackage{graphicx}
\usepackage{dcolumn}
\usepackage{bm}
\usepackage{amsmath,amssymb}
\usepackage{hyperref}
\usepackage{natbib}

\begin{document}

\preprint{APS/123-QED}

\title{Dynamical correlations in the electronic structure of BiFeO$_{3}$, as revealed by dynamical mean field theory} 

\author{Souvik Paul}
\email{paul.souvik@physics.uu.se}
\affiliation{Department of Physics and Astronomy, Materials Theory Division, Uppsala University, Box 516, SE-751 20 Uppsala, Sweden}
\author{Diana Iu\c{s}an}
\affiliation{Department of Physics and Astronomy, Materials Theory Division, Uppsala University, Box 516, SE-751 20 Uppsala, Sweden}
\author{P. Thunstr\"{o}m}
\affiliation{Department of Physics and Astronomy, Materials Theory Division, Uppsala University, Box 516, SE-751 20 Uppsala, Sweden}
\author{Y. O. Kvashnin}
\affiliation{Department of Physics and Astronomy, Materials Theory Division, Uppsala University, Box 516, SE-751 20 Uppsala, Sweden}
\author{Johan Hellsvik}
\affiliation{Department of Materials and Nano Physics, School of Information and Communication Technology, KTH Royal Institute of 
Technology, Electrum 229, SE-16440 Kista, Sweden}
\author{Manuel Pereiro}
\affiliation{Department of Physics and Astronomy, Materials Theory Division, Uppsala University, Box 516, SE-751 20 Uppsala, Sweden}
\author{A. Delin}
\affiliation{Department of Physics and Astronomy, Materials Theory Division, Uppsala University, Box 516, SE-751 20 Uppsala, Sweden}
\affiliation{Department of Materials and Nano Physics, School of Information and Communication Technology, KTH Royal Institute of 
Technology, Electrum 229, SE-16440 Kista, Sweden}
\affiliation{Swedish e-Science Research Center (SeRC), KTH Royal Institute of Technology, SE-10044 Stockholm, Sweden}
\author{Biplab Sanyal}
\affiliation{Department of Physics and Astronomy, Materials Theory Division, Uppsala University, Box 516, SE-751 20 Uppsala, Sweden}
\author{O. Eriksson}
\affiliation{Department of Physics and Astronomy, Materials Theory Division, Uppsala University, Box 516, SE-751 20 Uppsala, Sweden}

\date{\today}

\begin{abstract}
Using local density approximation plus dynamical mean-field theory (LDA+DMFT), we have computed the valence 
band photoelectron spectra of highly popular multiferroic BiFeO$_{3}$. Within DMFT, the local impurity 
problem is tackled by exact diagonalization (ED) solver. For comparison, we also present result from 
LDA+U approach, which is commonly used to compute physical properties of this compound. Our LDA+DMFT 
derived spectra match adequately with the experimental hard X-ray photoelectron spectroscopy (HAXPES) and resonant 
photoelectron spectroscopy (RPES) for Fe 3$d$ states, whereas the other theoretical method that we employed failed 
to capture the features of the measured spectra. Thus, our investigation shows the importance of accurately 
incorporating the dynamical aspects of electron-electron interaction among the Fe 3$d$ orbitals in calculations to produce 
the experimental excitation spectra, which establishes BiFeO$_{3}$ as a strongly correlated electron system. 
The LDA+DMFT derived density of states (DOSs) exhibit significant amount of Fe 3$d$ states at the energy 
of Bi lone-pairs, implying that the latter is not as alone as previously thought in the spectral scenario. 
Our study also demonstrates that the combination of orbital cross-sections for the constituent elements and 
broadening schemes for the calculated spectral function are pivotal to explain the detailed structures of 
the experimental spectra.
\end{abstract}

\pacs{71.27.+a, 71.20.Be, 79.60.-i}
\keywords{DMFT, strongly correlated multiferroic, theoretical spectroscopy}
\maketitle


\section{Introduction}

One of the primary interests in current materials research developed on multiferroics from the perspective of 
experimental and theoretical physics to understand the microscopic mechanisms controlling the observed properties 
\cite{ramesh2007}. Multiferroics combine two or more primary ferroic orders simultaneously in a single phase 
\cite{schmid1994}. In the conventional sense, now-a-days, multiferroics indicate coupling of ferroelectric 
polarization with any kind of magnetic order \cite{nicola_PT}. The magneto-electric (ME) coupling allows us to 
control the magnetization and the electric polarization with their corresponding conjugate fields. This functionality can 
be utilized to develop minuscule and energy-efficient devices for practical applications.

Extensive research in recent years on multiferroics has recognized BiFeO$_{3}$ as a promising candidate, 
which exhibits spontaneous ferroelectricity and magnetism at room temperature (ferroelectric T$_{C}$ $\sim$ 1103 K 
\cite{wang2003} and N\'{e}el temperature, T$_{N}\sim$ 643 K \cite{moreau1971}). It's stable polymorph at ambient 
pressure crystallize in rhombohedral perovskite structure\cite{michel1969}. The structure is characterized by 
counter-rotation of two oxygen octahedra (FeO$_{6}$) along the pseudocubic direction [111] and displacement of 
Fe atoms from the centre of octahedra along the same axis. The spontaneous ferroelectricity developed along 
[111] direction due to large displacive movement of Bi atoms relative to oxygen octahedra 
which is consistent with stereochemically active Bi lone-pairs \cite{neaton2005}. BiFeO$_{3}$ has complex 
magnetic structure whose origin is still under debate and investigations. The Fe atoms are known to couple 
ferromagnetically within (111) planes and antiferromagnetically between adjacent planes which 
results in G-type antiferromagnetic (AFM) order. An incommensurate cycloidal spin structure having wavelength of 
$\sim$620 $\AA$ is found to be superimposed with the antiparallel alignment of Fe moments \cite{spiral}. The symmetry 
of BiFeO$_{3}$ permits to develop small local ferromagnetic moment of Dzyaloshinskii-Moriya (DM) type by breaking 
the perfect antiparallel symmetry to a small canted one \cite{wferro,ederer2005}. However, 
the spin spiral configuration causes a cancellation of local ferromagnetic components over the volume encompassing 
the spiral which forces the average macroscopic magnetization to vanish. This restricts the material from exhibiting 
a linear ME effect \cite{kad2004}. Doping, application of large magnetic field and low dimensional structures can 
suppress the magnetic spin spiral and establish the material as a propitious candidate for ME device applications
\cite{popov1993,popov1994,kad2004,bras2009,bai2005}.  

Early attempts to compute the electronic structures of BiFeO$_{3}$ were made by first-principles based density 
functional theory (DFT) method where the exchange-correlation functional was treated under local spin 
density approximation (LSDA). This mean field-like approach failed to incorporate the correlation effect of 
localized Fe 3$d$ electrons and largely underestimated the measured insulating gap \cite{neaton2005,pio2005}. 
In an attempt to remedy this error, electronic structure calculations were performed by DFT(LSDA/GGA)+U method 
which treats the electronic correlation in a static mean-field fashion by adding an orbital-dependent on-site 
Hubbard U term to the Kohm-Sham Hamiltonian. The method splits the bands of localized 3$d$ electrons in lower 
(occupied) and upper (unoccupied) Hubbard bands and produces an insulating gap comparable to experimental 
observations \cite{neaton2005,pio2005,clark2007,shang2009,ju2009,goffinet2009,he2015,yaakob2015}. In DFT+U calculations 
\cite{neaton2005,clark2007,he2015}, the occupied part of the Fe 3$d$ density of states (DOSs) (from the Fermi level 
to $\sim$ -6 eV) agrees very poorly with the observation, e.g., RPES (resonant photoelectron spectroscopy) 
spectra for Fe 3$d$ states in Ref.~\onlinecite{maz2016}. The finding that DFT+U theory agrees poorly with 
measured spectra for complex oxides is also observed in Refs.~\onlinecite{oscar2012,patprl}. Studies based 
on hybrid functionals and self-interaction corrected theory also failed to improve the agreement with the measured 
electronic structure \cite{goffinet2009,die2011,yaakob2015}. 
In this regard, it should be noted that these methods worked well for BiFeO$_{3}$ in producing some 
experimental data like ferroelectric polarization, band gap, canting of the Fe moments etc. 
\cite{neaton2005,pio2005,clark2007,ederer2005}. 
However, failure of all these methods in reproducing the measured spectra simply reflects that the electronic 
structure of this compound is not well understood and requires a deeper theoretical analysis. 

In this paper, we make an overall comparison between experimental and theoretical spectra of BiFeO$_{3}$, using 
DFT+U method as well as more sophisticated many-body method based on dynamical mean-field theory (DMFT) 
\cite{geo_rev,kot_rev,held_rev}. We compare the calculated spectra with measured valence band spectra, 
as revealed by photoelectron spectroscopy (PES) \cite{maz2016}. DMFT method solves the Hubbard model by mapping it onto 
an Anderson Impurity Model (AIM) \cite{aim} where the impurity is embedded in a self-consistent medium and the 
interaction between impurity and its surrounding is described through hybridization function. In this method, 
information about the electronic excitations enters into the spectra of band structures via frequency-dependent 
self-energy, which contains knowledge of the many-body interactions. The DFT+U method is a static Hartree-Fock 
approximation to DMFT and it lacks the frequency dependency of the self-energy. This rationalizes the failure of 
DFT+U method in reproducing the spectroscopic features whereas ground state properties, at least in certain cases, can be 
calculated with acceptable accuracy. 

The DFT+DMFT approach has been applied successfully to several transition metal oxides (TMOs), e.g., the monoxides and the 
ABO$_{3}$ type perovskite \cite{r1,r2,r3,r4,r5,r6,r7,r8,r9,r10,r11,r12}. Recently, Shorikov {\it et al.} studied 
the metal-insulator transition (MIT) in BiFeO$_{3}$ using DFT+DMFT scheme employing continuous-time quantum Monte 
Carlo (CT-QMC) as an impurity solver \cite{bfodmft}. Their calculations in the paramagnetic phase, at high 
temperature (770 K) and at ambient pressure exhibited a band gap of 1.2 eV and thus illustrated that in contrast to 
DFT+U result, an insulating solution can be obtained in absence of long-range magnetic order, which is, in general, 
consistent with observations for transition metal oxides. However, direct comparison between calculated and 
measured spectra was not made in Ref.~\onlinecite{bfodmft}, since the focus of their investigation 
was rather on the pressure driven MIT. The fact that these calculations reproduced the observed MIT is clearly 
encouraging. 

Unfortunately, detailed investigations between experimental and calculated valence band spectra were not presented in 
Ref.~\onlinecite{bfodmft}. In fact, calculations based on CT-QMC may indeed be difficult to compare with the measured 
spectra due to the following reasons: i) the maximum entropy method (MEM) commonly used to 
analytically continue the Green's function to the real axis may produces spectral function that lacks sufficient
structure; ii) the precision of MEM decreases rapidly for energies away from the Fermi level which makes it more 
difficult to estimate the spectral function at higher binding energies. In fact, in Ref.\onlinecite{bfodmft}, data were 
shown only up to a range of 5 eV and hence, comparison to the measured valence band spectra was not possible.

The above discussion indicates that no attempt has been made from the theoretical community to explicitly analyze 
the experimental photoelectron spectra in BiFeO$_{3}$ and this has motivated us to perform detailed calculations 
of the electronic excitation spectra using the DFT+DMFT method in order to draw conclusions concerning the influence 
of electron correlations on the electronic structure of this much investigated material. To be specific, we compute 
the electronic spectra of BiFeO$_{3}$ with DFT+DMFT formalism and compare it to the DFT+U method and the 
experimental (hard X-ray photoelectron spectroscopy (HAXPES) and RPES) results\cite{maz2016}. The exact diagonalization 
(ED) scheme was chosen to solve the impurity problem within the DMFT approach, since in the past, it has been shown 
to give good spectroscopic data. The paper is organized as follows: section \ref{sec:comdet} encapsulates the details 
of the computational method used, in section \ref{sec:resdis}, we discuss about the spectral properties and 
finally we summarize our results in section \ref{sec:conc}. In Appendix \ref{sec:intpam}, we discuss the influence of 
different combinations of interaction parameters (U and J) on the spectra. 

\section{\label{sec:comdet}Computational details}
In this work, the rhombohedral crystal structure of BiFeO$_{3}$ with $R3c$ space group ($\#$ 161) was used for all 
the calculations \cite{str}. The electronic structures were calculated using full-potential linearized muffin-tin 
orbital (FP-LMTO) formalism under DFT as implemented in RSPt \cite{rspt1,rspt2}. The local (spin) density 
approximation (L(S)DA) as parameterized by von Barth-Hedin was used for the exchange-correlation part of the Kohn-Sham 
potential \cite{xc1,xc2}. The 5$d$, 6$s$ and 6$p$ orbitals of Bi; 3$d$, 4$s$ and 4$p$ orbitals of Fe and 2$s$ and 2$p$ 
orbitals of O were included in valence energy set for the construction of basis functions. The kinetic energy of 
basis functions in the interstitial (tails) were fixed at 0.3 Ry, -2.3 Ry and -1.5 Ry for $s$ and $p$ orbitals 
whereas the first two tails were considered for Bi 5$d$ states and for Fe 3$d$ orbitals, we use a tail with energy 
parameter set at -0.3 Ry. The $k$-points were distributed in a uniform $8\times8\times8$ Monkhorst-Pack grid centered 
at $\Gamma$-point and Brillouin zone integration was carried out using Fermi-Dirac smearing corresponding to T= 273 K. 
The Muffin-tin spheres of radii 2.26 a.u. for Bi, 1.88 a.u for Fe and 1.62 a.u. for O were chosen to match the charge 
density smoothly in the interstitial region. The on-site Coulomb interaction was parametrized by Hubbard U and Hund's 
exchange parameter J. In the literatures, U$_{eff}$ (=U-J) has been varied from 3 to 6 to match the experimental 
outcomes. Our main results are presented with U= 6 eV and J= 0.9 eV (U$_{eff}$= 5.1), which is close to the value used in 
CT-QMC based DMFT work (U= 6 eV and J= 0.93 eV) \cite{bfodmft}. Results with other U and J combinations are 
shown in Appendix \ref{sec:intpam}. The paramagnetic (PM) ED calculation was simulated using 20 auxiliary bath 
spin-orbitals and 10 Fe 3$d$ spin-orbitals. Among the 20 orbitals, 16 bath states are associated with two e$_{g}$ 
orbitals and the rest of them are associated with a$_{1g}$ orbitals. To reach convergence in the self-energy, 2632 
Matsubara frequencies were considered along the imaginary axis. The double counting problem was treat under the fully 
localized limit (FLL) approximation \cite{fll}. The Slater parameters F$^{2}$ and F$^{4}$ were obtained through fixed 
atomic ratio and found to be 0.58 and 0.36, respectively.

The theoretical photoelectron spectra were calculated using an effective single particle approximation for the 
photoelectrons together with an independent-scattering approximation for the final state wave function 
\cite{caroli1973,peter1974,mark1986}. Within these approximations, the angle-integrated 
photocurrent can be written as sum of the product of projected DOS and cross-section, where the summation 
spans over atom and orbital projected states. The availability of the self-consistent potentials from the DMFT method have 
been exploited to evaluate the cross-sections. The spectral lines were convoluted using a Lorentzian line shape function to 
include the finite lifetime of the excited states. We used Gaussian distribution to mimic the broadening due to 
instrumental resolution.

\section{\label{sec:resdis}Results and Discussion}
\subsection{\label{sec:dos} Density of states calculated using LSDA+U and LDA+DMFT methods}
\begin{figure}[h!]
\includegraphics[width=8.5cm]{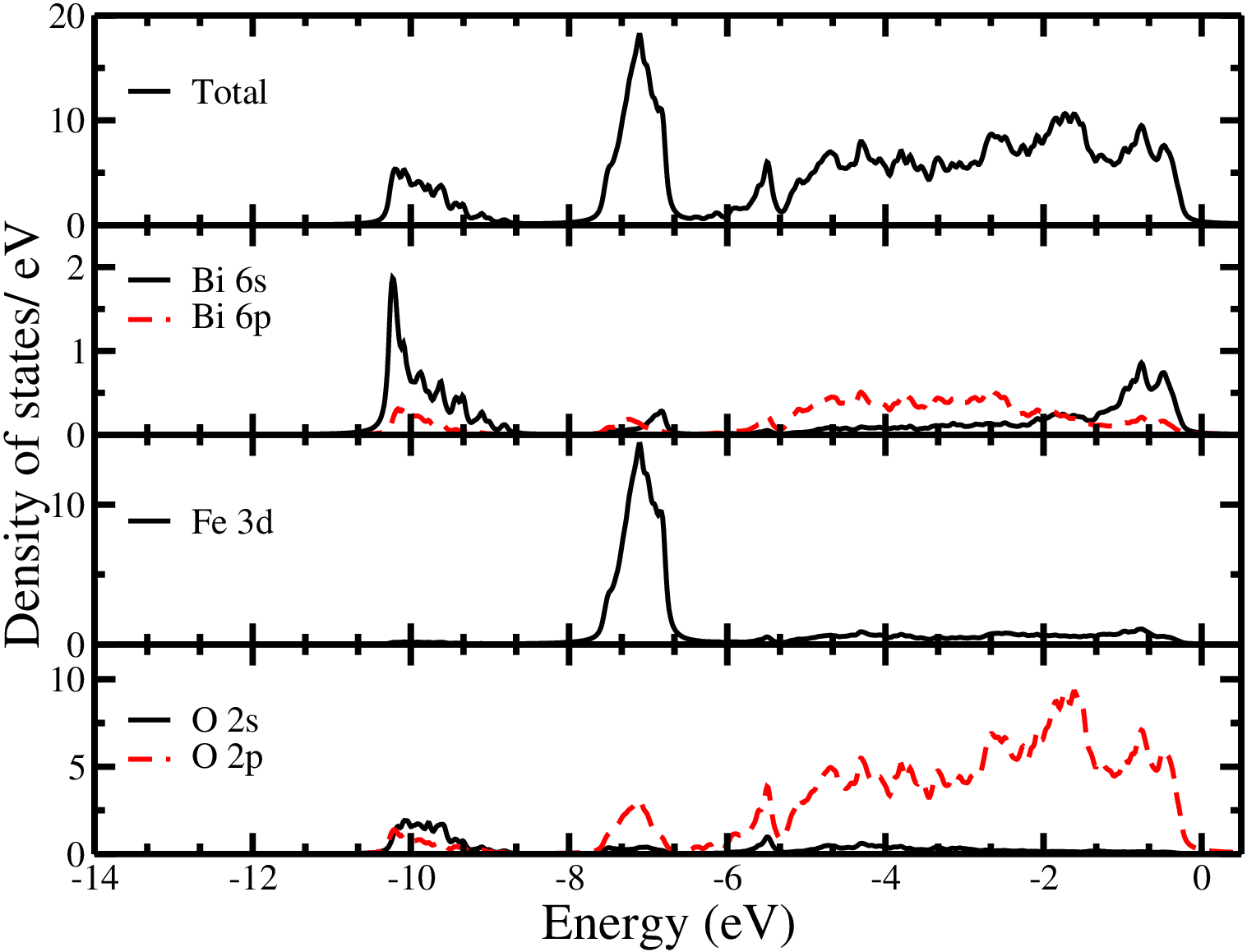}
\centering
\caption{Total and projected density of states of BiFeO$_{3}$ in the valence band, calculated using the LSDA+U method 
with U= 6 eV and J= 0.9 eV. The top of the valence band is set to zero.}
\label{fig:f1}
\end{figure}

\begin{figure}[h!]
\centering
\includegraphics[width=8.5cm]{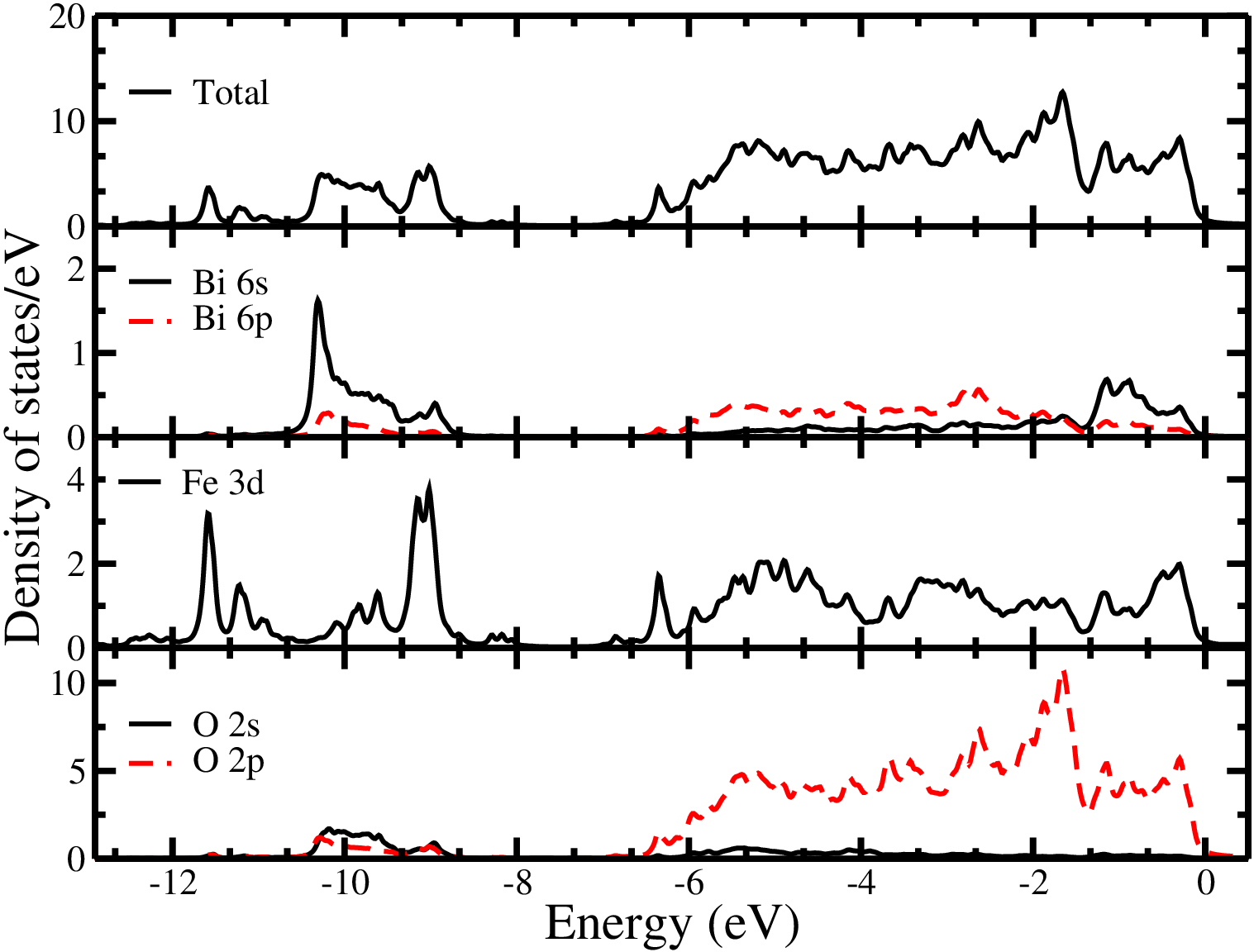}
\caption{Total and projected density of states of BiFeO$_{3}$ in the valence band, calculated using the LDA+DMFT method 
with U= 6 eV and J= 0.9 eV. The top of the valence band is set to zero.}
\label{fig:f2}
\end{figure}

In FIGs.\ref{fig:f1} and \ref{fig:f2}, we present total and orbital-projected valence band density of states (DOSs) 
evaluated using LSDA+U and LDA+DMFT methods, respectively. Each figure contains total DOSs 
in the top panel and Bi 6$s$, Bi 6$p$, Fe 3$d$, O 2$s$ and O 2$p$ projected DOSs (PDOSs) in the lower panels. 
The PDOSs of Bi 5$d$, Fe 4$s$ and Fe 4$p$ are extremely small and featureless over the considered energy range and 
therefore, exempted from the following discussions. 

A general inspection of the LSDA+U result (FIG. \ref{fig:f1}) indicates that the total DOSs from energy 0 eV to 
$\sim$-7 eV are caused by hybridization involving Fe, O and Bi states, which is consistent with previous reports. 
The Fe 3$d$ and O 2$p$ states interact strongly, whereas Bi 6$s$ and 6$p$ states couple weakly with the other two 
elements. The Bi states located around -10 eV predominantly have 6$s$ character and are attributed to stereochemically 
active lone-pairs. However, small amount of states from Bi 6$p$ and O 2$p$ orbitals are also found which compares 
well with the literatures. The DMFT result (FIG. \ref{fig:f2}) has several features that are similar to LSDA+U 
calculation, at least for Bi and O derived states. However, the spectral features of Fe 3$d$ states are 
markedly different, as can be seen by comparing FIGs.\ref{fig:f1} and \ref{fig:f2}. We will return to this discussion 
below, but first, we note that our calculated insulating gap using DMFT method in the PM phase is $\sim$1.22 eV, 
which tally well with Ref.~\onlinecite{bfodmft}. The value of the energy gap once again confirms that it is 
independent of long-range magnetic order. The conduction band minimum (CBM) in both the methods is characterized 
by states from the upper Hubbard band of Fe 3$d$ orbitals (data not shown). The valence band maximum (VBM) has leading 
contribution from Fe 3$d$ and O 2$p$ bands and trivial one from Bi 6$s$ and 6$p$ bands.

Comparison of FIGs.\ref{fig:f1} and \ref{fig:f2} shows that the total DOSs calculated from the two methods are 
markedly different. A detailed analysis reveals that the disagreement mainly occurs due to the differences in Fe 
3$d$ DOSs. The LSDA+U method pushes most of the Fe 3$d$ electrons into a region of very narrow energy range 
which is visible from sharp and high DOSs at around -7 eV with small hybridization with Bi and O states. The prominent 
Fe peak in the LSDA+U result is replaced by multiplet features in the DMFT calculation and the spectral weight is 
distributed over wider energy intervals. The 3$d$ DOSs are found to be significant in energy intervals from 
the Fermi level down to -6.5 eV and from -8.5 to -12 eV (see FIG. \ref{fig:f2}). The Fe 3$d$ states closer to the 
Fermi level are characterized by significant hybridization with O 2$p$ and Bi 6$p$ states, while the Fe 3$d$ states 
from -8.5 to -12 eV are less hybridized. However, intensity of the multiplet peaks at lower energy interval 
is significant and curiously their position overlap with Bi 6$s$ lone-pairs. Hence, 
the result of FIG. \ref{fig:f2} brings forth a picture where the Bi lone-pairs are not so lonely anymore and have 
significant spectral contribution from Fe 3$d$ multiplet structures around -10 eV. It is important to mention that, in our LDA+DMFT calculation, 
the electronic occupation of the Fe 3$d$ orbitals is 5.85, thus, referring to an electronic configuration closer to d$^6$.   

It is sufficient to compare the total DOSs and Fe 3$d$ PDOSs directly to measured HAXPES and RPES spectra, respectively, 
for demonstrating the superiority of the DMFT method and for interpretations of the measured spectra. However, for 
analyzing intricate details of the experimental spectra by comparing it to the theoretical results and thus, obtaining 
detailed understanding about the origin of all the spectral structures, one need to take the cross-section into account. 
We present such comparison and analysis in the next subsection. 

\subsection{\label{sec:spectra} Comparison between theoretical and experimental spectra 
and analysis of the spectral behavior}
\begin{figure}[h!]
\centering
\includegraphics[width=8.5cm]{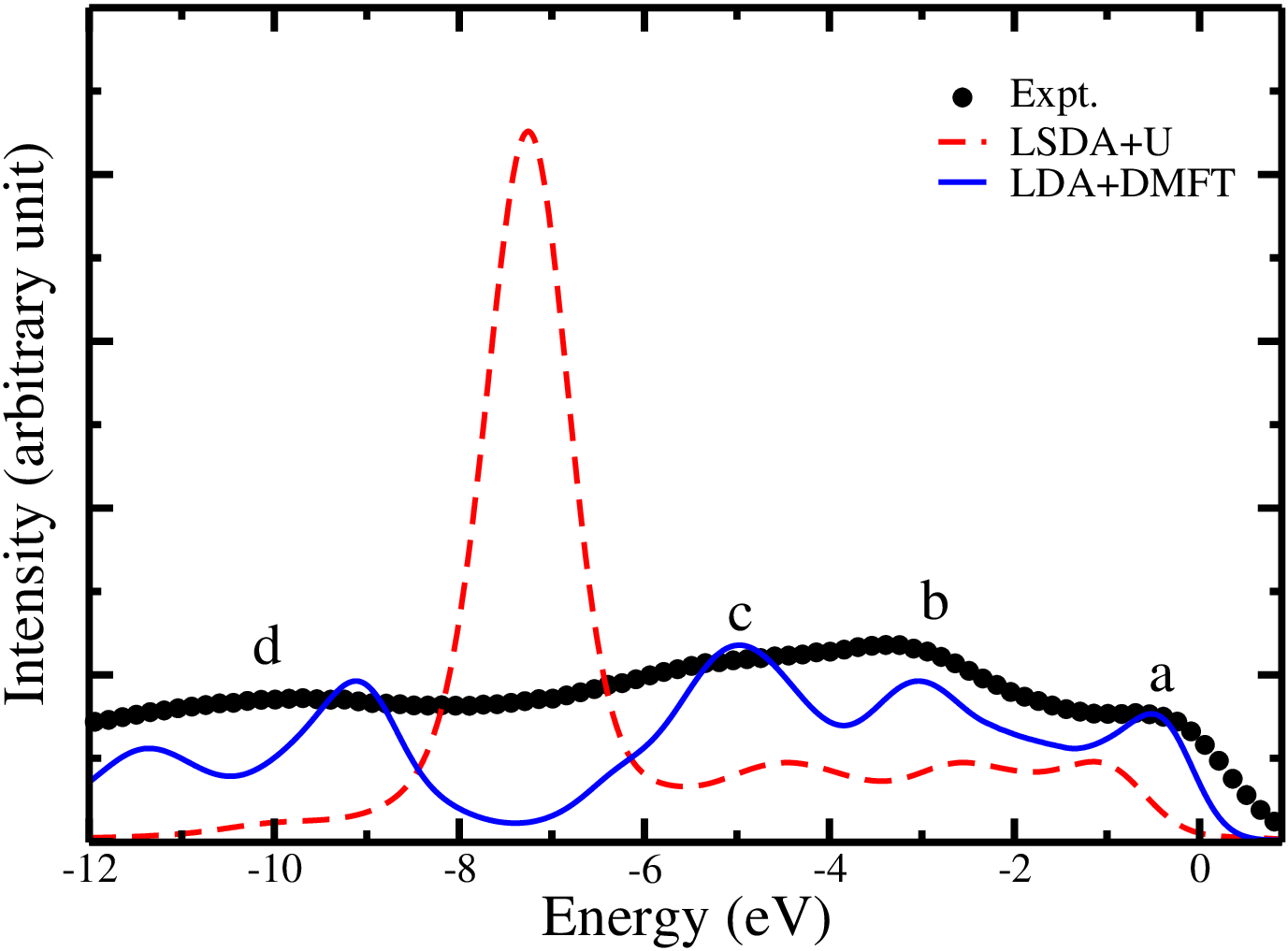}
\caption{A comparison between the experimental RPES spectra (Ref. \onlinecite{maz2016}) and the theoretical spectra for Fe 3$d$ 
states of BiFeO$_{3}$ using LSDA+U and DMFT methods. The theoretical calculations are performed using U= 6 eV and J= 0.9 eV. The experimental data 
are shifted to align with the calculated valence band edge (see text).}
\label{fig:f3}
\end{figure}

\begin{figure}[h!]
\centering
\includegraphics[width=8.5cm]{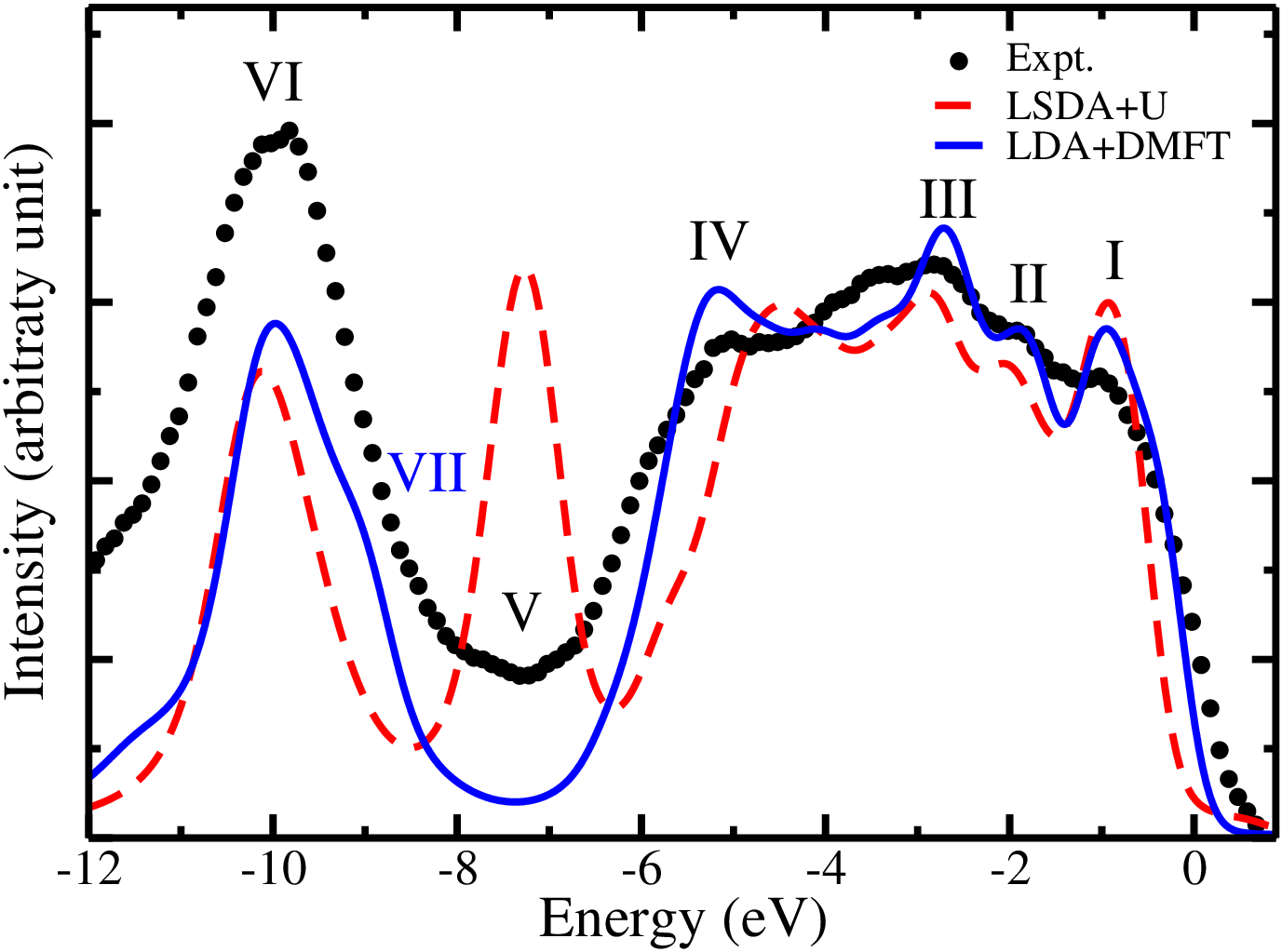}
\caption{A comparison between the experimental HAXPES spectra (Ref. \onlinecite{maz2016}) and the theoretical spectra of BiFeO$_{3}$ 
using LSDA+U and DMFT methods. The theoretical calculations were performed using U= 6 eV and J= 0.9 eV. The experimental data 
are shifted to align with the calculated valence band edge (see text).}
\label{fig:f4}
\end{figure}

In FIGs.\ref{fig:f3} and \ref{fig:f4}, we compare the calculated Fe 3$d$ projected and total electronic spectra from
LSDA+U and LDA+DMFT approaches to experimental RPES and HAXPES spectra, respectively \cite{maz2016}. Note that in FIG. 
\ref{fig:f4}, cross-sections of all the valence states are taken into consideration. The experiment was performed at 
room temperature and HAXPES and RPES data were collected at photon energy of 2 keV and at photon energy corresponding 
to L$_{3}$ absorption peak of Fe, respectively. The latter photon energy produces resonant measurement that primarily 
signifies Fe 3$d$ states. It is noticeable in the experimental spectra published by Mazumdar {\it et al.} 
\cite{maz2016} that the position of the Fermi level is not located at the valence band maximum, which often is due to 
defects in the sample. To remove the ambiguity in Fermi level, we shifted the experimental spectra to match the 
valence band onset. 

\begin{figure*}[t]
\centering
\includegraphics[width=11.5cm]{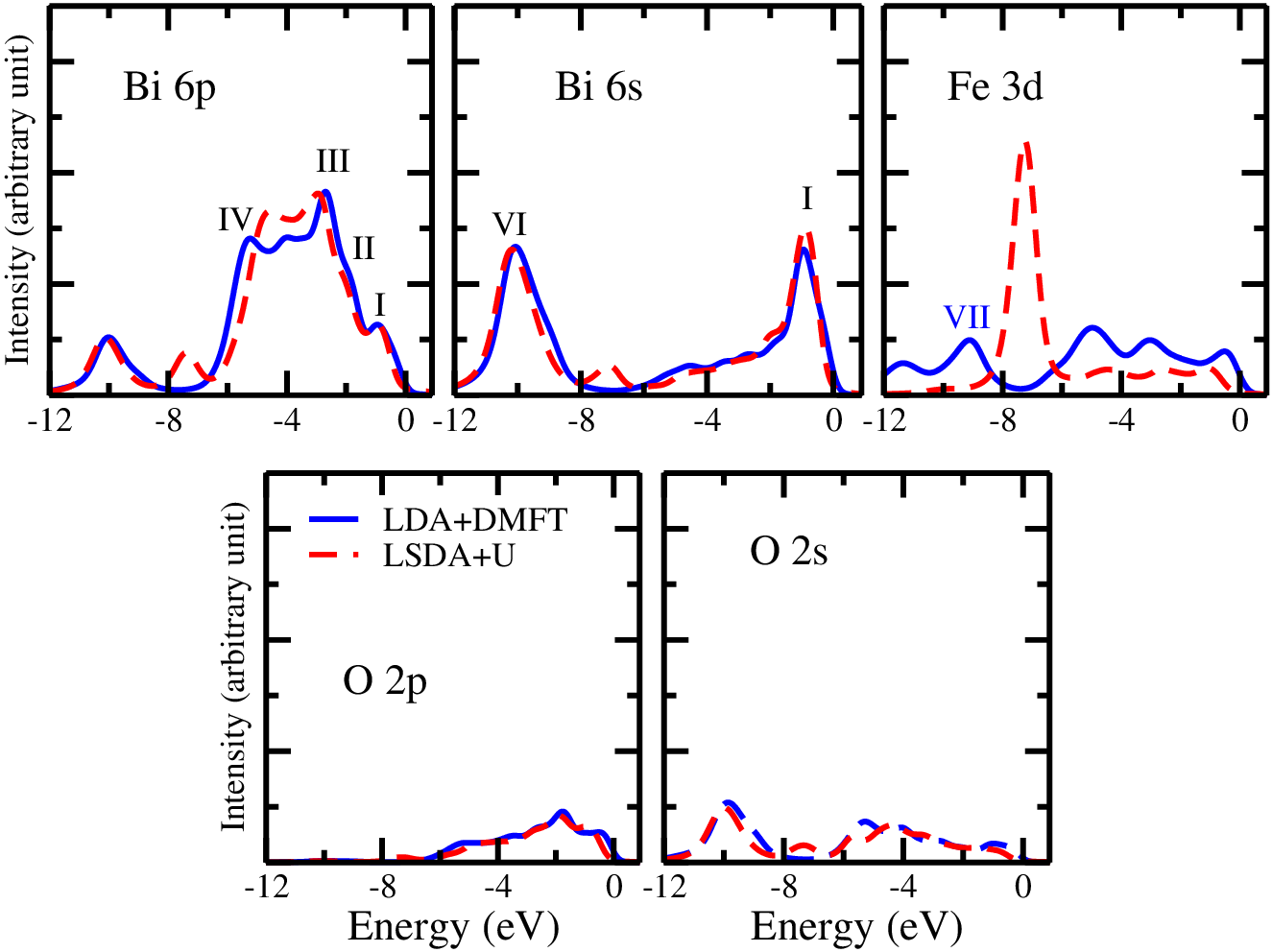}
\caption{Atom and orbital projected spectral states of BiFeO$_{3}$ from LDA+U and DMFT methods. The calculations are done with Coulomb 
interaction parameters U= 6 eV and J= 0.9 eV. All the spectra are normalized in intensity with respect to the Bi 5$d$ state 
(data not shown).}
\label{fig:f5}
\end{figure*}

In FIG. \ref{fig:f3}, the RPES spectra show a three-peak structure, denoted by $a$, $b$ and $c$, between 0 eV to -6 eV, 
where the third peak ($c$) is less conspicuous compared to the others. There is also a broad and weaker spectral contribution, 
denoted by $d$, that is centered around -10 eV. A general comparison illustrates that the LSDA+U result reproduces the 
experimental spectra with poor accuracy. The intensity of the spectra computed from LSDA+U method, in the energy range 
between 0 eV and -6 eV, is very small and shows marginal variations; then it rapidly enhanced to a sharp peak at around -7 eV 
and thenceforth, it vanishes quickly with further increase in binding energy. Overall, this does not represent the general 
features as captured in the experiment. The LSDA+U result generates a three-peak structures below -6 eV, as also found in 
the RPES spectra. However, the position and intensity of LSDA+U derived peaks do not coincide with the experimental result. 
Furthermore, the large peak at around -7 eV is entirely missing from the experimental observation. It is pellucid from the 
fact that the RPES spectra exhibit a dip in the intensity pattern at the position of that theoretical peak. The discrepancy 
also persists at higher binding energies, since no significant amount of Fe 3$d$ states are found in FIG. \ref{fig:f1} 
that match the measured hump centered around -10 eV in the resonant spectra. 

On the other hand, DMFT result conforms better with the RPES spectra regarding the positions of all the peaks 
and the width of the spectra. In theory, we found significant amount of intensity in the entire energy range in accordance with the experimental data. 
Note that measured intensity at binding energies higher than 12 eV is caused by secondary emission which forms a general 
background to the spectra. The intensity of the spectra for the DMFT method is found to increase gradually from peak $a$ to $c$, 
in agreement with the experiment. Also, non-negligible contribution to the spectra at higher binding energies up to 12 eV 
is perfectly captured by the theory and as discussed above this is due 
to 3$d$ multiplets at lower energy interval (see FIG. \ref{fig:f2}). The measured spectra in FIG. \ref{fig:f3} 
have a small dip at around -7.5 eV, which is also present in the theory, albeit here the dip is more pronounced. Although the 
reason for this disagreement is a matter of debate, but we conclude that the overall features are well captured. The aforesaid 
discourse indicates that the correlation effects among the Fe 3$d$ electrons are important and properly considered within the 
DMFT calculation.

In FIG. \ref{fig:f4}, the valence band HAXPES spectra show a sharp peak near the Fermi level at -0.9 eV (feature I), 
followed by two more closely positioned peaks at -1.85 eV and -2.7 eV, denoted by features II and III, respectively. A distant 
fourth peak is located at -5.2 eV (feature IV), separated from peak III by a region having little variation in intensity. 
Furthermore, around -10 eV, we observe an extended crest (feature VI) succeed by a valley (feature V). A careful inspection 
shows that at low and high binding energies, both the theoretical methods generate similar results which are more or less 
consistent with the HAXPES spectra. However, in the intermediate and high binding energy range, the LSDA+U result deviates from 
experimental result significantly, whereas the DMFT result matches with the experimental spectra. 

One a more detailed level, we note that the structures of our theoretical spectra for the two methods down to -3 eV match 
more or less well with the three peaks of the HAXPES data both in terms of position and intensity. Thereafter, the LSDA+U 
result starts to depart from the experimental data and hence, the first countable difference arises as the position of the 
fourth peak (IV) is underestimated by $\sim$0.7 eV in the spectra calculated using LSDA+U method, whereas the DMFT derived spectra 
better captures the characteristic of peak IV. The region with minimal variation between peak III and IV of the experimental 
spectra is nicely captured in the DMFT result, showing a smooth variation of intensity with energy. The second and most 
noticeable difference occurs due to the sharp peak in LSDA+U result, that is located around -7 eV. This originates, as discussed 
above, from the high intensity peak related to Fe 3$d$ states. In contrast to the LSDA+U result, we observe a valley in the DMFT 
spectra at that energy which is consistent with the measured HAXPES spectra. Around -10 eV, qualitatively both the theoretical 
methods produce a broad pattern in accordance with the HAXPES spectra, although discernible difference in 
intensity between theoretical and experimental spectra is observed. This difference is due to the fact that we do not include background signals in our 
calculation which is present in the experiment. We have chosen not to include the background since its absence does not alter 
our physical reasoning of different features in the spectra. 
The finer details of peak VI shows a weak structure in the HAXPES measurements that seem to be composed by two overlapping peaks 
(FIG. \ref{fig:f4}). This is not captured by the LSDA+U calculation that has a single smooth feature, nor by the DMFT calculation 
which has more structure compared to the previous method. However, the DMFT derived spectra exhibit a shoulder at feature VII, 
instead of two-peak structure. 

In order to analyze the elemental contributions to the theoretical spectra, we present orbital projected states 
of Bi 6$s$, Bi 6$p$, Fe 3$d$, O 2$s$, and O 2$p$ in FIG. \ref{fig:f5}. Both the DMFT and the LSDA+U results are presented. Although 
the DOSs for Bi in FIG. \ref{fig:f2} are small relative to Fe and O throughout the energy range, due to higher cross-section, 
the element has largest spectral weight and mostly decides the structures of the spectra. On the other hand, due to lower 
cross-section, the spectral intensity of Fe and O states have less significant contributions to the spectra. 
The first peak (I) has biggest contribution from Bi 6$s$ states and moderate amount of contribution from Bi 6$p$ states. 
From there on, the intensity of Bi 6$s$ states decreases promptly, becomes comparable to the intensity of Fe and O states 
and less important for features up to -6 eV. In contrast, the intensity of Bi 6$p$ states increase rapidly and dominate the 
intermediate peaks II, III and IV. The Fe 3$d$, O 2$s$ and O 2$p$ states offer a small weight to the spectra 
throughout the energy range. The extended feature VI is governed by Bi 6$s$ states but have influences from Bi 6$p$ and O 2$s$ 
states and from the DMFT calculations, also the Fe 3$d$ states. In all the figures, spectral characteristic computed using 
the LSDA+U method deviates slightly from the DMFT results, except for Fe 3$d$ states, where the difference 
is most significant. The spectral properties are mostly affected by the changes in Fe 3$d$ spectra and similarities near the two extreme 
energy limits are primarily controlled by Bi 6$s$ and Bi 6$p$ spectra. 

\section{\label{sec:conc}Conclusion}
In this paper, we resolve the discrepancies in the electronic excitation spectra of extensively investigated multiferroic 
BiFeO$_{3}$ between experiment and LSDA+U method with improved results from LDA+DMFT method. Our calculations demonstrate 
that an accurate description of the electronic correlation among the Fe 3$d$ electrons is necessary to obtain good agreement 
with the experiment. The LSDA+U method offers poor description of the electronic structure due to its limited treatment of 
correlation effects. For the two methods, we observe that the changes in DOSs are most prominent for Fe 3$d$ states, whereas 
it is trivial for the other two constituent elements. The high-intensity narrow peak in Fe 3$d$ DOSs obtained under LSDA+U method, 
redistributed under LDA+DMFT approach and generates multiplet structures that spread over wider energy regions. 
The Fe 3$d$ multiplets at lower energies appear at the position of Bi 6$s$ lone-pairs states, signifying hybridization 
between those orbitals, which is completely absent in 
LSDA+U picture. The reorganized Fe 3$d$ DOSs in LDA+DMFT method when multiplied with cross-section capture the features 
of the experimental RPES spectra, whereas the other method drastically fails to even produce the general features of the 
measured spectra. Similarly, on the overall energy range, the features produce by LDA+DMFT method agrees 
well with the experimental HAXPES measurement. Inspection of element resolved LDA+DMFT derived spectra reveals that Bi mostly 
governs the behavior of the spectra and the other two element have small contributions. The same investigation for the two 
theoretical methods illustrates that similarities in the spectra are due to Bi derived states and changes are caused by 
Fe 3$d$ states. Summarizing, we produce theoretical valence band spectra for BiFeO$_{3}$ with LDA+DMFT method which nicely matches 
with all the experimental features and shows that dynamical correlation is important for this multiferroic material.

\begin{acknowledgments}
The simulations have been performed on supercomputer provided by National Supercomputer Centre and 
PDC Center for High Performance Computing under the project of Swedish National Infrastructure for 
Computing (SNIC). We also acknowledged support from the Swedish Research Council (VR), the KAW 
foundation (grants  2012.0031 and 2013.0020) and eSSENCE. The authors would like to thank Ronny Knut 
and Olof Karis for valuable discussions regarding the measured HAXPES and RPES spectra. The author 
S.P thanks S. K. Panda for discussions at the initial stage of simulation.   
\end{acknowledgments}

\appendix
\section{\label{sec:intpam} Influence of the interaction parameters on the spectra}


\begin{figure}[h!]
\centering
\includegraphics[width=8.5cm]{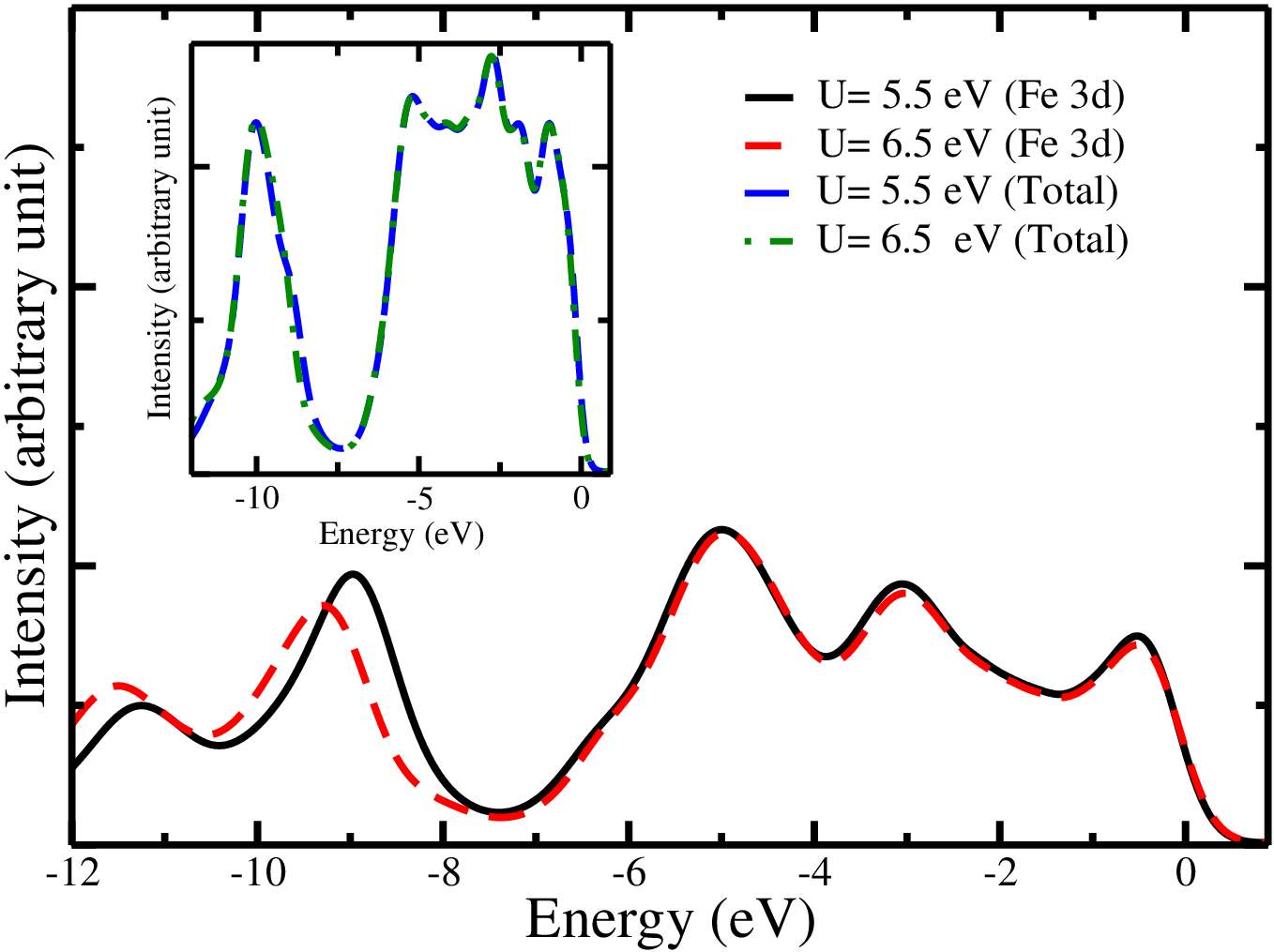}
\caption{Theoretical spectra of Fe 3$d$ states calculated under LDA+DMFT method 
for two different U values, keeping J fixed to 0.9 eV. The inset shows total spectra 
for the same U values.}
\label{fig:f6}
\end{figure}

It is interesting to examine the impact of the Coulomb interaction parameter U on the spectra. For this, we have performed 
LDA+DMFT calculations for two different values of U, below and above 6 eV, with fixed J= 0.9 eV. In FIG. \ref{fig:f6}, 
we present total and Fe 3$d$ projected spectra of such calculations. On an overall basis, features in the total spectra remain 
static to the variation of Hubbard parameter U, whereas some changes are observed in the Fe 3$d$ spectra. The basic features 
of the two curves for Fe 3$d$ in FIG. \ref{fig:f6} are similar to one in FIG. \ref{fig:f3}; however, features at higher binding 
energies move sightly away from the Fermi level when U is increased. The strong hybridization of Fe 3$d$ orbitals with 
the bath orbitals (O 2$p$) between 0 and -6 eV allows to overcome the effect of variation in U and consequently no changes are 
observed in the spectra. On the contrary, due to extremely low 
degree of hybridization, Fe 3$d$ states at higher binding energies behave close to atomic-like and result in 
the observed behavior. We observe minimal changes when all these spectra are compared with J= 1.0 eV, keeping U fixed (data not shown). 
From this discussion, we can conclude that the choices of interaction parameters do not alter the physical apprehension of the 
spectral properties in this material as long as their values are plausible. 

Our calculated electronic spectra using DFMT method at photon energy of 6 keV differ 
significantly at higher binding energies in terms of intensity with the 
measured spectra in Ref.~\onlinecite{maz2016}. Recently, with the aid of DMFT+GW calculations in NiO, Panda {\it et al.} \cite{panda2016} 
showed that the effect of non-local correlation on the delocalized states is important to explain the structures 
of excitation spectra at high photon energy. Based on their observation, we come to the resolution that our local correlation based 
DMFT result is not sufficient to produce spectra at such high energy for BiFeO$_{3}$ and we leave the investigation 
by including non-local correlations within the calculations for future publication. 


\begin{thebibliography}{53}%
\makeatletter
\providecommand \@ifxundefined [1]{%
 \@ifx{#1\undefined}
}%
\providecommand \@ifnum [1]{%
 \ifnum #1\expandafter \@firstoftwo
 \else \expandafter \@secondoftwo
 \fi
}%
\providecommand \@ifx [1]{%
 \ifx #1\expandafter \@firstoftwo
 \else \expandafter \@secondoftwo
 \fi
}%
\providecommand \natexlab [1]{#1}%
\providecommand \enquote  [1]{``#1''}%
\providecommand \bibnamefont  [1]{#1}%
\providecommand \bibfnamefont [1]{#1}%
\providecommand \citenamefont [1]{#1}%
\providecommand \href@noop [0]{\@secondoftwo}%
\providecommand \href [0]{\begingroup \@sanitize@url \@href}%
\providecommand \@href[1]{\@@startlink{#1}\@@href}%
\providecommand \@@href[1]{\endgroup#1\@@endlink}%
\providecommand \@sanitize@url [0]{\catcode `\\12\catcode `\$12\catcode
  `\&12\catcode `\#12\catcode `\^12\catcode `\_12\catcode `\%12\relax}%
\providecommand \@@startlink[1]{}%
\providecommand \@@endlink[0]{}%
\providecommand \url  [0]{\begingroup\@sanitize@url \@url }%
\providecommand \@url [1]{\endgroup\@href {#1}{\urlprefix }}%
\providecommand \urlprefix  [0]{URL }%
\providecommand \Eprint [0]{\href }%
\providecommand \doibase [0]{http://dx.doi.org/}%
\providecommand \selectlanguage [0]{\@gobble}%
\providecommand \bibinfo  [0]{\@secondoftwo}%
\providecommand \bibfield  [0]{\@secondoftwo}%
\providecommand \translation [1]{[#1]}%
\providecommand \BibitemOpen [0]{}%
\providecommand \bibitemStop [0]{}%
\providecommand \bibitemNoStop [0]{.\EOS\space}%
\providecommand \EOS [0]{\spacefactor3000\relax}%
\providecommand \BibitemShut  [1]{\csname bibitem#1\endcsname}%
\let\auto@bib@innerbib\@empty
\bibitem [{\citenamefont {Ramesh}\ and\ \citenamefont
  {Spaldin}(2007)}]{ramesh2007}%
  \BibitemOpen
  \bibfield  {author} {\bibinfo {author} {\bibfnamefont {R.}~\bibnamefont
  {Ramesh}}\ and\ \bibinfo {author} {\bibfnamefont {N.~A.}\ \bibnamefont
  {Spaldin}},\ }\href {\doibase 10.1038/nmat1805} {\bibfield  {journal}
  {\bibinfo  {journal} {Nat Mater}\ }\textbf {\bibinfo {volume} {6}},\ \bibinfo
  {pages} {21} (\bibinfo {year} {2007})}\BibitemShut {NoStop}%
\bibitem [{\citenamefont {Schmid}(1994)}]{schmid1994}%
  \BibitemOpen
  \bibfield  {author} {\bibinfo {author} {\bibfnamefont {H.}~\bibnamefont
  {Schmid}},\ }\href {\doibase 10.1080/00150199408245120} {\bibfield  {journal}
  {\bibinfo  {journal} {Ferroelectrics}\ }\textbf {\bibinfo {volume} {162}},\
  \bibinfo {pages} {317} (\bibinfo {year} {1994})}\BibitemShut {NoStop}%
\bibitem [{\citenamefont {Spaldin}\ \emph {et~al.}(2010)\citenamefont
  {Spaldin}, \citenamefont {Cheong},\ and\ \citenamefont {Ramesh}}]{nicola_PT}%
  \BibitemOpen
  \bibfield  {author} {\bibinfo {author} {\bibfnamefont {N.~A.}\ \bibnamefont
  {Spaldin}}, \bibinfo {author} {\bibfnamefont {S.-W.}\ \bibnamefont {Cheong}},
  \ and\ \bibinfo {author} {\bibfnamefont {R.}~\bibnamefont {Ramesh}},\ }\href
  {\doibase 10.1063/1.3502547} {\bibfield  {journal} {\bibinfo  {journal}
  {Phys. Today}\ }\textbf {\bibinfo {volume} {63}},\ \bibinfo {pages} {38}
  (\bibinfo {year} {2010})}\BibitemShut {NoStop}%
\bibitem [{\citenamefont {Wang}\ \emph {et~al.}(2003)\citenamefont {Wang},
  \citenamefont {Neaton}, \citenamefont {Zheng}, \citenamefont {Nagarajan},
  \citenamefont {Ogale}, \citenamefont {Liu}, \citenamefont {Viehland},
  \citenamefont {Vaithyanathan}, \citenamefont {Schlom}, \citenamefont
  {Waghmare}, \citenamefont {Spaldin}, \citenamefont {Rabe}, \citenamefont
  {Wuttig},\ and\ \citenamefont {Ramesh}}]{wang2003}%
  \BibitemOpen
  \bibfield  {author} {\bibinfo {author} {\bibfnamefont {J.}~\bibnamefont
  {Wang}}, \bibinfo {author} {\bibfnamefont {J.~B.}\ \bibnamefont {Neaton}},
  \bibinfo {author} {\bibfnamefont {H.}~\bibnamefont {Zheng}}, \bibinfo
  {author} {\bibfnamefont {V.}~\bibnamefont {Nagarajan}}, \bibinfo {author}
  {\bibfnamefont {S.~B.}\ \bibnamefont {Ogale}}, \bibinfo {author}
  {\bibfnamefont {B.}~\bibnamefont {Liu}}, \bibinfo {author} {\bibfnamefont
  {D.}~\bibnamefont {Viehland}}, \bibinfo {author} {\bibfnamefont
  {V.}~\bibnamefont {Vaithyanathan}}, \bibinfo {author} {\bibfnamefont {D.~G.}\
  \bibnamefont {Schlom}}, \bibinfo {author} {\bibfnamefont {U.~V.}\
  \bibnamefont {Waghmare}}, \bibinfo {author} {\bibfnamefont {N.~A.}\
  \bibnamefont {Spaldin}}, \bibinfo {author} {\bibfnamefont {K.~M.}\
  \bibnamefont {Rabe}}, \bibinfo {author} {\bibfnamefont {M.}~\bibnamefont
  {Wuttig}}, \ and\ \bibinfo {author} {\bibfnamefont {R.}~\bibnamefont
  {Ramesh}},\ }\href {\doibase 10.1126/science.1080615} {\bibfield  {journal}
  {\bibinfo  {journal} {Science}\ }\textbf {\bibinfo {volume} {299}},\ \bibinfo
  {pages} {1719} (\bibinfo {year} {2003})}\BibitemShut {NoStop}%
\bibitem [{\citenamefont {Moreau}\ \emph {et~al.}(1971)\citenamefont {Moreau},
  \citenamefont {Michel}, \citenamefont {Gerson},\ and\ \citenamefont
  {James}}]{moreau1971}%
  \BibitemOpen
  \bibfield  {author} {\bibinfo {author} {\bibfnamefont {J.}~\bibnamefont
  {Moreau}}, \bibinfo {author} {\bibfnamefont {C.}~\bibnamefont {Michel}},
  \bibinfo {author} {\bibfnamefont {R.}~\bibnamefont {Gerson}}, \ and\ \bibinfo
  {author} {\bibfnamefont {W.}~\bibnamefont {James}},\ }\href {\doibase
  10.1016/S0022-3697(71)80189-0} {\bibfield  {journal} {\bibinfo  {journal} {J.
  Phys. Chem. Solids}\ }\textbf {\bibinfo {volume} {32}},\ \bibinfo {pages}
  {1315} (\bibinfo {year} {1971})}\BibitemShut {NoStop}%
\bibitem [{\citenamefont {Michel}\ \emph {et~al.}(1969)\citenamefont {Michel},
  \citenamefont {Moreau}, \citenamefont {Achenbach}, \citenamefont {Gerson},\
  and\ \citenamefont {James}}]{michel1969}%
  \BibitemOpen
  \bibfield  {author} {\bibinfo {author} {\bibfnamefont {C.}~\bibnamefont
  {Michel}}, \bibinfo {author} {\bibfnamefont {J.-M.}\ \bibnamefont {Moreau}},
  \bibinfo {author} {\bibfnamefont {G.~D.}\ \bibnamefont {Achenbach}}, \bibinfo
  {author} {\bibfnamefont {R.}~\bibnamefont {Gerson}}, \ and\ \bibinfo {author}
  {\bibfnamefont {W.}~\bibnamefont {James}},\ }\href {\doibase
  10.1016/0038-1098(69)90597-3} {\bibfield  {journal} {\bibinfo  {journal}
  {Solid State Commun.}\ }\textbf {\bibinfo {volume} {7}},\ \bibinfo {pages}
  {701} (\bibinfo {year} {1969})}\BibitemShut {NoStop}%
\bibitem [{\citenamefont {Neaton}\ \emph {et~al.}(2005)\citenamefont {Neaton},
  \citenamefont {Ederer}, \citenamefont {Waghmare}, \citenamefont {Spaldin},\
  and\ \citenamefont {Rabe}}]{neaton2005}%
  \BibitemOpen
  \bibfield  {author} {\bibinfo {author} {\bibfnamefont {J.~B.}\ \bibnamefont
  {Neaton}}, \bibinfo {author} {\bibfnamefont {C.}~\bibnamefont {Ederer}},
  \bibinfo {author} {\bibfnamefont {U.~V.}\ \bibnamefont {Waghmare}}, \bibinfo
  {author} {\bibfnamefont {N.~A.}\ \bibnamefont {Spaldin}}, \ and\ \bibinfo
  {author} {\bibfnamefont {K.~M.}\ \bibnamefont {Rabe}},\ }\href {\doibase
  10.1103/PhysRevB.71.014113} {\bibfield  {journal} {\bibinfo  {journal} {Phys.
  Rev. B}\ }\textbf {\bibinfo {volume} {71}},\ \bibinfo {pages} {014113}
  (\bibinfo {year} {2005})}\BibitemShut {NoStop}%
\bibitem [{\citenamefont {Sosnowska}\ \emph {et~al.}(1982)\citenamefont
  {Sosnowska}, \citenamefont {Neumaier},\ and\ \citenamefont
  {Steichele}}]{spiral}%
  \BibitemOpen
  \bibfield  {author} {\bibinfo {author} {\bibfnamefont {I.}~\bibnamefont
  {Sosnowska}}, \bibinfo {author} {\bibfnamefont {T.~P.}\ \bibnamefont
  {Neumaier}}, \ and\ \bibinfo {author} {\bibfnamefont {E.}~\bibnamefont
  {Steichele}},\ }\href {http://stacks.iop.org/0022-3719/15/i=23/a=020}
  {\bibfield  {journal} {\bibinfo  {journal} {J. Phys. C: Solid State Phys.}\
  }\textbf {\bibinfo {volume} {15}},\ \bibinfo {pages} {4835} (\bibinfo {year}
  {1982})}\BibitemShut {NoStop}%
\bibitem [{\citenamefont {Zhang}\ \emph {et~al.}(2005)\citenamefont {Zhang},
  \citenamefont {Lu}, \citenamefont {Wu}, \citenamefont {Chen},\ and\
  \citenamefont {Ming}}]{wferro}%
  \BibitemOpen
  \bibfield  {author} {\bibinfo {author} {\bibfnamefont {S.~T.}\ \bibnamefont
  {Zhang}}, \bibinfo {author} {\bibfnamefont {M.~H.}\ \bibnamefont {Lu}},
  \bibinfo {author} {\bibfnamefont {D.}~\bibnamefont {Wu}}, \bibinfo {author}
  {\bibfnamefont {Y.~F.}\ \bibnamefont {Chen}}, \ and\ \bibinfo {author}
  {\bibfnamefont {N.~B.}\ \bibnamefont {Ming}},\ }\href
  {http://scitation.aip.org/content/aip/journal/apl/87/26/10.1063/1.2147719}
  {\bibfield  {journal} {\bibinfo  {journal} {Appl. Phys. Lett.}\ }\textbf
  {\bibinfo {volume} {87}},\ \bibinfo {eid} {262907} (\bibinfo {year}
  {2005})}\BibitemShut {NoStop}%
\bibitem [{\citenamefont {Ederer}\ and\ \citenamefont
  {Spaldin}(2005)}]{ederer2005}%
  \BibitemOpen
  \bibfield  {author} {\bibinfo {author} {\bibfnamefont {C.}~\bibnamefont
  {Ederer}}\ and\ \bibinfo {author} {\bibfnamefont {N.~A.}\ \bibnamefont
  {Spaldin}},\ }\href {\doibase 10.1103/PhysRevB.71.060401} {\bibfield
  {journal} {\bibinfo  {journal} {Phys. Rev. B}\ }\textbf {\bibinfo {volume}
  {71}},\ \bibinfo {pages} {060401} (\bibinfo {year} {2005})}\BibitemShut
  {NoStop}%
\bibitem [{\citenamefont {Kadomtseva}\ \emph {et~al.}(2004)\citenamefont
  {Kadomtseva}, \citenamefont {Zvezdin}, \citenamefont {Popov}, \citenamefont
  {Pyatakov},\ and\ \citenamefont {Vorob'ev}}]{kad2004}%
  \BibitemOpen
  \bibfield  {author} {\bibinfo {author} {\bibfnamefont {A.~M.}\ \bibnamefont
  {Kadomtseva}}, \bibinfo {author} {\bibfnamefont {A.~K.}\ \bibnamefont
  {Zvezdin}}, \bibinfo {author} {\bibfnamefont {Y.~F.}\ \bibnamefont {Popov}},
  \bibinfo {author} {\bibfnamefont {A.~P.}\ \bibnamefont {Pyatakov}}, \ and\
  \bibinfo {author} {\bibfnamefont {G.~P.}\ \bibnamefont {Vorob'ev}},\ }\href
  {\doibase 10.1134/1.1787107} {\bibfield  {journal} {\bibinfo  {journal}
  {JETP. Lett.}\ }\textbf {\bibinfo {volume} {79}},\ \bibinfo {pages} {571}
  (\bibinfo {year} {2004})}\BibitemShut {NoStop}%
\bibitem [{\citenamefont {Popov}\ \emph {et~al.}(1993)\citenamefont {Popov},
  \citenamefont {Zvezdin}, \citenamefont {Vorob{\rq}ev}, \citenamefont
  {Kadomtseva}, \citenamefont {Murashev},\ and\ \citenamefont
  {Rakov}}]{popov1993}%
  \BibitemOpen
  \bibfield  {author} {\bibinfo {author} {\bibfnamefont {Y.~F.}\ \bibnamefont
  {Popov}}, \bibinfo {author} {\bibfnamefont {A.~K.}\ \bibnamefont {Zvezdin}},
  \bibinfo {author} {\bibfnamefont {G.~P.}\ \bibnamefont {Vorob{\rq}ev}},
  \bibinfo {author} {\bibfnamefont {A.~M.}\ \bibnamefont {Kadomtseva}},
  \bibinfo {author} {\bibfnamefont {V.~A.}\ \bibnamefont {Murashev}}, \ and\
  \bibinfo {author} {\bibfnamefont {D.~N.}\ \bibnamefont {Rakov}},\ }\href@noop
  {} {\bibfield  {journal} {\bibinfo  {journal} {JETP Lett.}\ }\textbf
  {\bibinfo {volume} {57}},\ \bibinfo {pages} {69} (\bibinfo {year}
  {1993})}\BibitemShut {NoStop}%
\bibitem [{\citenamefont {Popov}\ \emph {et~al.}(1994)\citenamefont {Popov},
  \citenamefont {Kadomtseva}, \citenamefont {Vorob{\rq}ev},\ and\ \citenamefont
  {Zvezdin}}]{popov1994}%
  \BibitemOpen
  \bibfield  {author} {\bibinfo {author} {\bibfnamefont {Y.~F.}\ \bibnamefont
  {Popov}}, \bibinfo {author} {\bibfnamefont {A.~M.}\ \bibnamefont
  {Kadomtseva}}, \bibinfo {author} {\bibfnamefont {G.~P.}\ \bibnamefont
  {Vorob{\rq}ev}}, \ and\ \bibinfo {author} {\bibfnamefont {A.~K.}\
  \bibnamefont {Zvezdin}},\ }\href@noop {} {\bibfield  {journal} {\bibinfo
  {journal} {Ferroelectrics}\ }\textbf {\bibinfo {volume} {162}},\ \bibinfo
  {pages} {135} (\bibinfo {year} {1994})}\BibitemShut {NoStop}%
\bibitem [{\citenamefont {{Le Bras}}\ \emph {et~al.}(2009)\citenamefont {{Le
  Bras}}, \citenamefont {Colson}, \citenamefont {Forget}, \citenamefont
  {Genand-Riondet}, \citenamefont {Tourbot},\ and\ \citenamefont
  {Bonville}}]{bras2009}%
  \BibitemOpen
  \bibfield  {author} {\bibinfo {author} {\bibfnamefont {G.}~\bibnamefont {{Le
  Bras}}}, \bibinfo {author} {\bibfnamefont {D.}~\bibnamefont {Colson}},
  \bibinfo {author} {\bibfnamefont {A.}~\bibnamefont {Forget}}, \bibinfo
  {author} {\bibfnamefont {N.}~\bibnamefont {Genand-Riondet}}, \bibinfo
  {author} {\bibfnamefont {R.}~\bibnamefont {Tourbot}}, \ and\ \bibinfo
  {author} {\bibfnamefont {P.}~\bibnamefont {Bonville}},\ }\href {\doibase
  10.1103/PhysRevB.80.134417} {\bibfield  {journal} {\bibinfo  {journal} {Phys.
  Rev. B}\ }\textbf {\bibinfo {volume} {80}},\ \bibinfo {pages} {134417}
  (\bibinfo {year} {2009})}\BibitemShut {NoStop}%
\bibitem [{\citenamefont {Bai}\ \emph {et~al.}(2005)\citenamefont {Bai},
  \citenamefont {Wang}, \citenamefont {Wuttig}, \citenamefont {Li},
  \citenamefont {Wang}, \citenamefont {Pyatakov}, \citenamefont {Zvezdin},
  \citenamefont {Cross},\ and\ \citenamefont {Viehland}}]{bai2005}%
  \BibitemOpen
  \bibfield  {author} {\bibinfo {author} {\bibfnamefont {F.}~\bibnamefont
  {Bai}}, \bibinfo {author} {\bibfnamefont {J.}~\bibnamefont {Wang}}, \bibinfo
  {author} {\bibfnamefont {M.}~\bibnamefont {Wuttig}}, \bibinfo {author}
  {\bibfnamefont {J.}~\bibnamefont {Li}}, \bibinfo {author} {\bibfnamefont
  {N.}~\bibnamefont {Wang}}, \bibinfo {author} {\bibfnamefont {A.~P.}\
  \bibnamefont {Pyatakov}}, \bibinfo {author} {\bibfnamefont {A.~K.}\
  \bibnamefont {Zvezdin}}, \bibinfo {author} {\bibfnamefont {L.~E.}\
  \bibnamefont {Cross}}, \ and\ \bibinfo {author} {\bibfnamefont
  {D.}~\bibnamefont {Viehland}},\ }\href
  {http://scitation.aip.org/content/aip/journal/apl/86/3/10.1063/1.1851612}
  {\bibfield  {journal} {\bibinfo  {journal} {Appl. Phys. Lett.}\ }\textbf
  {\bibinfo {volume} {86}},\ \bibinfo {eid} {032511} (\bibinfo {year}
  {2005})}\BibitemShut {NoStop}%
\bibitem [{\citenamefont {Baettig}\ \emph {et~al.}(2005)\citenamefont
  {Baettig}, \citenamefont {Ederer},\ and\ \citenamefont {Spaldin}}]{pio2005}%
  \BibitemOpen
  \bibfield  {author} {\bibinfo {author} {\bibfnamefont {P.}~\bibnamefont
  {Baettig}}, \bibinfo {author} {\bibfnamefont {C.}~\bibnamefont {Ederer}}, \
  and\ \bibinfo {author} {\bibfnamefont {N.~A.}\ \bibnamefont {Spaldin}},\
  }\href {\doibase 10.1103/PhysRevB.72.214105} {\bibfield  {journal} {\bibinfo
  {journal} {Phys. Rev. B}\ }\textbf {\bibinfo {volume} {72}},\ \bibinfo
  {pages} {214105} (\bibinfo {year} {2005})}\BibitemShut {NoStop}%
\bibitem [{\citenamefont {Clark}\ and\ \citenamefont
  {Robertson}(2007)}]{clark2007}%
  \BibitemOpen
  \bibfield  {author} {\bibinfo {author} {\bibfnamefont {S.~J.}\ \bibnamefont
  {Clark}}\ and\ \bibinfo {author} {\bibfnamefont {J.}~\bibnamefont
  {Robertson}},\ }\href
  {http://scitation.aip.org/content/aip/journal/apl/90/13/10.1063/1.2716868}
  {\bibfield  {journal} {\bibinfo  {journal} {Appl. Phys. Lett.}\ }\textbf
  {\bibinfo {volume} {90}},\ \bibinfo {eid} {132903} (\bibinfo {year}
  {2007})}\BibitemShut {NoStop}%
\bibitem [{\citenamefont {Shang}\ \emph {et~al.}(2009)\citenamefont {Shang},
  \citenamefont {Sheng}, \citenamefont {Wang}, \citenamefont {Chen},\ and\
  \citenamefont {Liu}}]{shang2009}%
  \BibitemOpen
  \bibfield  {author} {\bibinfo {author} {\bibfnamefont {S.~L.}\ \bibnamefont
  {Shang}}, \bibinfo {author} {\bibfnamefont {G.}~\bibnamefont {Sheng}},
  \bibinfo {author} {\bibfnamefont {Y.}~\bibnamefont {Wang}}, \bibinfo {author}
  {\bibfnamefont {L.~Q.}\ \bibnamefont {Chen}}, \ and\ \bibinfo {author}
  {\bibfnamefont {Z.~K.}\ \bibnamefont {Liu}},\ }\href {\doibase
  10.1103/PhysRevB.80.052102} {\bibfield  {journal} {\bibinfo  {journal} {Phys.
  Rev. B}\ }\textbf {\bibinfo {volume} {80}},\ \bibinfo {pages} {052102}
  (\bibinfo {year} {2009})}\BibitemShut {NoStop}%
\bibitem [{\citenamefont {Ju}\ \emph {et~al.}(2009)\citenamefont {Ju},
  \citenamefont {Cai},\ and\ \citenamefont {Guo}}]{ju2009}%
  \BibitemOpen
  \bibfield  {author} {\bibinfo {author} {\bibfnamefont {S.}~\bibnamefont
  {Ju}}, \bibinfo {author} {\bibfnamefont {T.-Y.}\ \bibnamefont {Cai}}, \ and\
  \bibinfo {author} {\bibfnamefont {G.-Y.}\ \bibnamefont {Guo}},\ }\href
  {http://scitation.aip.org/content/aip/journal/jcp/130/21/10.1063/1.3146796}
  {\bibfield  {journal} {\bibinfo  {journal} {J. Chem. Phys.}\ }\textbf
  {\bibinfo {volume} {130}},\ \bibinfo {eid} {214708} (\bibinfo {year}
  {2009})}\BibitemShut {NoStop}%
\bibitem [{\citenamefont {Goffinet}\ \emph {et~al.}(2009)\citenamefont
  {Goffinet}, \citenamefont {Hermet}, \citenamefont {Bilc},\ and\ \citenamefont
  {Ghosez}}]{goffinet2009}%
  \BibitemOpen
  \bibfield  {author} {\bibinfo {author} {\bibfnamefont {M.}~\bibnamefont
  {Goffinet}}, \bibinfo {author} {\bibfnamefont {P.}~\bibnamefont {Hermet}},
  \bibinfo {author} {\bibfnamefont {D.~I.}\ \bibnamefont {Bilc}}, \ and\
  \bibinfo {author} {\bibfnamefont {P.}~\bibnamefont {Ghosez}},\ }\href
  {\doibase 10.1103/PhysRevB.79.014403} {\bibfield  {journal} {\bibinfo
  {journal} {Phys. Rev. B}\ }\textbf {\bibinfo {volume} {79}},\ \bibinfo
  {pages} {014403} (\bibinfo {year} {2009})}\BibitemShut {NoStop}%
\bibitem [{\citenamefont {He}\ \emph {et~al.}(2015)\citenamefont {He},
  \citenamefont {Ma}, \citenamefont {Sun}, \citenamefont {Sa},\ and\
  \citenamefont {Wu}}]{he2015}%
  \BibitemOpen
  \bibfield  {author} {\bibinfo {author} {\bibfnamefont {C.}~\bibnamefont
  {He}}, \bibinfo {author} {\bibfnamefont {Z.-J.}\ \bibnamefont {Ma}}, \bibinfo
  {author} {\bibfnamefont {B.-Z.}\ \bibnamefont {Sun}}, \bibinfo {author}
  {\bibfnamefont {R.-J.}\ \bibnamefont {Sa}}, \ and\ \bibinfo {author}
  {\bibfnamefont {K.}~\bibnamefont {Wu}},\ }\href {\doibase
  10.1016/j.jallcom.2014.11.062} {\bibfield  {journal} {\bibinfo  {journal} {J.
  Alloys Compd.}\ }\textbf {\bibinfo {volume} {623}},\ \bibinfo {pages} {393}
  (\bibinfo {year} {2015})}\BibitemShut {NoStop}%
\bibitem [{\citenamefont {Yaakob}\ \emph {et~al.}(2015)\citenamefont {Yaakob},
  \citenamefont {Taib}, \citenamefont {Lu}, \citenamefont {Hassan},\ and\
  \citenamefont {Yahya}}]{yaakob2015}%
  \BibitemOpen
  \bibfield  {author} {\bibinfo {author} {\bibfnamefont {M.~K.}\ \bibnamefont
  {Yaakob}}, \bibinfo {author} {\bibfnamefont {M.~F.~M.}\ \bibnamefont {Taib}},
  \bibinfo {author} {\bibfnamefont {L.}~\bibnamefont {Lu}}, \bibinfo {author}
  {\bibfnamefont {O.~H.}\ \bibnamefont {Hassan}}, \ and\ \bibinfo {author}
  {\bibfnamefont {M.~Z.~A.}\ \bibnamefont {Yahya}},\ }\href
  {http://stacks.iop.org/2053-1591/2/i=11/a=116101} {\bibfield  {journal}
  {\bibinfo  {journal} {Mater. Res. Express}\ }\textbf {\bibinfo {volume}
  {2}},\ \bibinfo {pages} {116101} (\bibinfo {year} {2015})}\BibitemShut
  {NoStop}%
\bibitem [{\citenamefont {Mazumdar}\ \emph {et~al.}(2016)\citenamefont
  {Mazumdar}, \citenamefont {Knut}, \citenamefont {Th{\"o}le}, \citenamefont
  {Gorgoi}, \citenamefont {Faleev}, \citenamefont {Mryasov}, \citenamefont
  {Shelke}, \citenamefont {Ederer}, \citenamefont {Spaldin}, \citenamefont
  {Gupta},\ and\ \citenamefont {Karis}}]{maz2016}%
  \BibitemOpen
  \bibfield  {author} {\bibinfo {author} {\bibfnamefont {D.}~\bibnamefont
  {Mazumdar}}, \bibinfo {author} {\bibfnamefont {R.}~\bibnamefont {Knut}},
  \bibinfo {author} {\bibfnamefont {F.}~\bibnamefont {Th{\"o}le}}, \bibinfo
  {author} {\bibfnamefont {M.}~\bibnamefont {Gorgoi}}, \bibinfo {author}
  {\bibfnamefont {S.}~\bibnamefont {Faleev}}, \bibinfo {author} {\bibfnamefont
  {O.}~\bibnamefont {Mryasov}}, \bibinfo {author} {\bibfnamefont
  {V.}~\bibnamefont {Shelke}}, \bibinfo {author} {\bibfnamefont
  {C.}~\bibnamefont {Ederer}}, \bibinfo {author} {\bibfnamefont
  {N.}~\bibnamefont {Spaldin}}, \bibinfo {author} {\bibfnamefont
  {A.}~\bibnamefont {Gupta}}, \ and\ \bibinfo {author} {\bibfnamefont
  {O.}~\bibnamefont {Karis}},\ }\href {\doibase 10.1016/j.elspec.2015.10.002}
  {\bibfield  {journal} {\bibinfo  {journal} {J. Electron Spectrosc. Relat.
  Phenom.}\ }\textbf {\bibinfo {volume} {208}},\ \bibinfo {pages} {63}
  (\bibinfo {year} {2016})},\ \bibinfo {note} {special Issue: Electronic
  structure and function from state-of-the-art spectroscopy and
  theory}\BibitemShut {NoStop}%
\bibitem [{\citenamefont {Gr{\aa}n{\"a}s}\ \emph {et~al.}(2012)\citenamefont
  {Gr{\aa}n{\"a}s}, \citenamefont {Marco}, \citenamefont {Thunstr{\"o}m},
  \citenamefont {Nordstr{\"o}m}, \citenamefont {Eriksson}, \citenamefont
  {Bj{\"o}rkman},\ and\ \citenamefont {Wills}}]{oscar2012}%
  \BibitemOpen
  \bibfield  {author} {\bibinfo {author} {\bibfnamefont {O.}~\bibnamefont
  {Gr{\aa}n{\"a}s}}, \bibinfo {author} {\bibfnamefont {I.~D.}\ \bibnamefont
  {Marco}}, \bibinfo {author} {\bibfnamefont {P.}~\bibnamefont
  {Thunstr{\"o}m}}, \bibinfo {author} {\bibfnamefont {L.}~\bibnamefont
  {Nordstr{\"o}m}}, \bibinfo {author} {\bibfnamefont {O.}~\bibnamefont
  {Eriksson}}, \bibinfo {author} {\bibfnamefont {T.}~\bibnamefont
  {Bj{\"o}rkman}}, \ and\ \bibinfo {author} {\bibfnamefont {J.}~\bibnamefont
  {Wills}},\ }\href {\doibase 10.1016/j.commatsci.2011.11.032} {\bibfield
  {journal} {\bibinfo  {journal} {Comput. Mater. Sci.}\ }\textbf {\bibinfo
  {volume} {55}},\ \bibinfo {pages} {295} (\bibinfo {year} {2012})}\BibitemShut
  {NoStop}%
\bibitem [{\citenamefont {Thunstr{\"o}m}\ \emph {et~al.}(2012)\citenamefont
  {Thunstr{\"o}m}, \citenamefont {{Di Marco}},\ and\ \citenamefont
  {Eriksson}}]{patprl}%
  \BibitemOpen
  \bibfield  {author} {\bibinfo {author} {\bibfnamefont {P.}~\bibnamefont
  {Thunstr{\"o}m}}, \bibinfo {author} {\bibfnamefont {I.}~\bibnamefont {{Di
  Marco}}}, \ and\ \bibinfo {author} {\bibfnamefont {O.}~\bibnamefont
  {Eriksson}},\ }\href {\doibase 10.1103/PhysRevLett.109.186401} {\bibfield
  {journal} {\bibinfo  {journal} {Phys. Rev. Lett.}\ }\textbf {\bibinfo
  {volume} {109}},\ \bibinfo {pages} {186401} (\bibinfo {year}
  {2012})}\BibitemShut {NoStop}%
\bibitem [{\citenamefont {Di{\'e}guez}\ \emph {et~al.}(2011)\citenamefont
  {Di{\'e}guez}, \citenamefont {Gonz{\'a}lez-V{\'a}zquez}, \citenamefont
  {Wojde{\l}},\ and\ \citenamefont {{\'I}{\~n}iguez}}]{die2011}%
  \BibitemOpen
  \bibfield  {author} {\bibinfo {author} {\bibfnamefont {O.}~\bibnamefont
  {Di{\'e}guez}}, \bibinfo {author} {\bibfnamefont {O.~E.}\ \bibnamefont
  {Gonz{\'a}lez-V{\'a}zquez}}, \bibinfo {author} {\bibfnamefont {J.~C.}\
  \bibnamefont {Wojde{\l}}}, \ and\ \bibinfo {author} {\bibfnamefont
  {J.}~\bibnamefont {{\'I}{\~n}iguez}},\ }\href {\doibase
  10.1103/PhysRevB.83.094105} {\bibfield  {journal} {\bibinfo  {journal} {Phys.
  Rev. B}\ }\textbf {\bibinfo {volume} {83}},\ \bibinfo {pages} {094105}
  (\bibinfo {year} {2011})}\BibitemShut {NoStop}%
\bibitem [{\citenamefont {Georges}\ \emph {et~al.}(1996)\citenamefont
  {Georges}, \citenamefont {Kotliar}, \citenamefont {Krauth},\ and\
  \citenamefont {Rozenberg}}]{geo_rev}%
  \BibitemOpen
  \bibfield  {author} {\bibinfo {author} {\bibfnamefont {A.}~\bibnamefont
  {Georges}}, \bibinfo {author} {\bibfnamefont {G.}~\bibnamefont {Kotliar}},
  \bibinfo {author} {\bibfnamefont {W.}~\bibnamefont {Krauth}}, \ and\ \bibinfo
  {author} {\bibfnamefont {M.~J.}\ \bibnamefont {Rozenberg}},\ }\href {\doibase
  10.1103/RevModPhys.68.13} {\bibfield  {journal} {\bibinfo  {journal} {Rev.
  Mod. Phys.}\ }\textbf {\bibinfo {volume} {68}},\ \bibinfo {pages} {13}
  (\bibinfo {year} {1996})}\BibitemShut {NoStop}%
\bibitem [{\citenamefont {Kotliar}\ \emph {et~al.}(2006)\citenamefont
  {Kotliar}, \citenamefont {Savrasov}, \citenamefont {Haule}, \citenamefont
  {Oudovenko}, \citenamefont {Parcollet},\ and\ \citenamefont
  {Marianetti}}]{kot_rev}%
  \BibitemOpen
  \bibfield  {author} {\bibinfo {author} {\bibfnamefont {G.}~\bibnamefont
  {Kotliar}}, \bibinfo {author} {\bibfnamefont {S.~Y.}\ \bibnamefont
  {Savrasov}}, \bibinfo {author} {\bibfnamefont {K.}~\bibnamefont {Haule}},
  \bibinfo {author} {\bibfnamefont {V.~S.}\ \bibnamefont {Oudovenko}}, \bibinfo
  {author} {\bibfnamefont {O.}~\bibnamefont {Parcollet}}, \ and\ \bibinfo
  {author} {\bibfnamefont {C.~A.}\ \bibnamefont {Marianetti}},\ }\href
  {\doibase 10.1103/RevModPhys.78.865} {\bibfield  {journal} {\bibinfo
  {journal} {Rev. Mod. Phys.}\ }\textbf {\bibinfo {volume} {78}},\ \bibinfo
  {pages} {865} (\bibinfo {year} {2006})}\BibitemShut {NoStop}%
\bibitem [{\citenamefont {Held}(2007)}]{held_rev}%
  \BibitemOpen
  \bibfield  {author} {\bibinfo {author} {\bibfnamefont {K.}~\bibnamefont
  {Held}},\ }\href {\doibase 10.1080/00018730701619647} {\bibfield  {journal}
  {\bibinfo  {journal} {Advances in Physics}\ }\textbf {\bibinfo {volume}
  {56}},\ \bibinfo {pages} {829} (\bibinfo {year} {2007})}\BibitemShut
  {NoStop}%
\bibitem [{\citenamefont {Anderson}(1961)}]{aim}%
  \BibitemOpen
  \bibfield  {author} {\bibinfo {author} {\bibfnamefont {P.~W.}\ \bibnamefont
  {Anderson}},\ }\href {\doibase 10.1103/PhysRev.124.41} {\bibfield  {journal}
  {\bibinfo  {journal} {Phys. Rev.}\ }\textbf {\bibinfo {volume} {124}},\
  \bibinfo {pages} {41} (\bibinfo {year} {1961})}\BibitemShut {NoStop}%
\bibitem [{\citenamefont {Pavarini}\ \emph {et~al.}(2004)\citenamefont
  {Pavarini}, \citenamefont {Biermann}, \citenamefont {Poteryaev},
  \citenamefont {Lichtenstein}, \citenamefont {Georges},\ and\ \citenamefont
  {Andersen}}]{r1}%
  \BibitemOpen
  \bibfield  {author} {\bibinfo {author} {\bibfnamefont {E.}~\bibnamefont
  {Pavarini}}, \bibinfo {author} {\bibfnamefont {S.}~\bibnamefont {Biermann}},
  \bibinfo {author} {\bibfnamefont {A.}~\bibnamefont {Poteryaev}}, \bibinfo
  {author} {\bibfnamefont {A.~I.}\ \bibnamefont {Lichtenstein}}, \bibinfo
  {author} {\bibfnamefont {A.}~\bibnamefont {Georges}}, \ and\ \bibinfo
  {author} {\bibfnamefont {O.~K.}\ \bibnamefont {Andersen}},\ }\href {\doibase
  10.1103/PhysRevLett.92.176403} {\bibfield  {journal} {\bibinfo  {journal}
  {Phys. Rev. Lett.}\ }\textbf {\bibinfo {volume} {92}},\ \bibinfo {pages}
  {176403} (\bibinfo {year} {2004})}\BibitemShut {NoStop}%
\bibitem [{\citenamefont {Pavarini}\ \emph {et~al.}(2005)\citenamefont
  {Pavarini}, \citenamefont {Yamasaki}, \citenamefont {Nuss},\ and\
  \citenamefont {Andersen}}]{r2}%
  \BibitemOpen
  \bibfield  {author} {\bibinfo {author} {\bibfnamefont {E.}~\bibnamefont
  {Pavarini}}, \bibinfo {author} {\bibfnamefont {A.}~\bibnamefont {Yamasaki}},
  \bibinfo {author} {\bibfnamefont {J.}~\bibnamefont {Nuss}}, \ and\ \bibinfo
  {author} {\bibfnamefont {O.~K.}\ \bibnamefont {Andersen}},\ }\href
  {http://stacks.iop.org/1367-2630/7/i=1/a=188} {\bibfield  {journal} {\bibinfo
   {journal} {New Journal of Physics}\ }\textbf {\bibinfo {volume} {7}},\
  \bibinfo {pages} {188} (\bibinfo {year} {2005})}\BibitemShut {NoStop}%
\bibitem [{\citenamefont {Solovyev}(2008)}]{r3}%
  \BibitemOpen
  \bibfield  {author} {\bibinfo {author} {\bibfnamefont {I.~V.}\ \bibnamefont
  {Solovyev}},\ }\href {http://stacks.iop.org/0953-8984/20/i=29/a=293201}
  {\bibfield  {journal} {\bibinfo  {journal} {Journal of Physics: Condensed
  Matter}\ }\textbf {\bibinfo {volume} {20}},\ \bibinfo {pages} {293201}
  (\bibinfo {year} {2008})}\BibitemShut {NoStop}%
\bibitem [{\citenamefont {{De Raychaudhury}}\ \emph {et~al.}(2007)\citenamefont
  {{De Raychaudhury}}, \citenamefont {Pavarini},\ and\ \citenamefont
  {Andersen}}]{r4}%
  \BibitemOpen
  \bibfield  {author} {\bibinfo {author} {\bibfnamefont {M.}~\bibnamefont {{De
  Raychaudhury}}}, \bibinfo {author} {\bibfnamefont {E.}~\bibnamefont
  {Pavarini}}, \ and\ \bibinfo {author} {\bibfnamefont {O.~K.}\ \bibnamefont
  {Andersen}},\ }\href {\doibase 10.1103/PhysRevLett.99.126402} {\bibfield
  {journal} {\bibinfo  {journal} {Phys. Rev. Lett.}\ }\textbf {\bibinfo
  {volume} {99}},\ \bibinfo {pages} {126402} (\bibinfo {year}
  {2007})}\BibitemShut {NoStop}%
\bibitem [{\citenamefont {Nekrasov}\ \emph {et~al.}(2005)\citenamefont
  {Nekrasov}, \citenamefont {Keller}, \citenamefont {Kondakov}, \citenamefont
  {Kozhevnikov}, \citenamefont {Pruschke}, \citenamefont {Held}, \citenamefont
  {Vollhardt},\ and\ \citenamefont {Anisimov}}]{r5}%
  \BibitemOpen
  \bibfield  {author} {\bibinfo {author} {\bibfnamefont {I.~A.}\ \bibnamefont
  {Nekrasov}}, \bibinfo {author} {\bibfnamefont {G.}~\bibnamefont {Keller}},
  \bibinfo {author} {\bibfnamefont {D.~E.}\ \bibnamefont {Kondakov}}, \bibinfo
  {author} {\bibfnamefont {A.~V.}\ \bibnamefont {Kozhevnikov}}, \bibinfo
  {author} {\bibfnamefont {T.}~\bibnamefont {Pruschke}}, \bibinfo {author}
  {\bibfnamefont {K.}~\bibnamefont {Held}}, \bibinfo {author} {\bibfnamefont
  {D.}~\bibnamefont {Vollhardt}}, \ and\ \bibinfo {author} {\bibfnamefont
  {V.~I.}\ \bibnamefont {Anisimov}},\ }\href {\doibase
  10.1103/PhysRevB.72.155106} {\bibfield  {journal} {\bibinfo  {journal} {Phys.
  Rev. B}\ }\textbf {\bibinfo {volume} {72}},\ \bibinfo {pages} {155106}
  (\bibinfo {year} {2005})}\BibitemShut {NoStop}%
\bibitem [{\citenamefont {Ohta}\ \emph {et~al.}(2012)\citenamefont {Ohta},
  \citenamefont {Cohen}, \citenamefont {Hirose}, \citenamefont {Haule},
  \citenamefont {Shimizu},\ and\ \citenamefont {Ohishi}}]{r6}%
  \BibitemOpen
  \bibfield  {author} {\bibinfo {author} {\bibfnamefont {K.}~\bibnamefont
  {Ohta}}, \bibinfo {author} {\bibfnamefont {R.~E.}\ \bibnamefont {Cohen}},
  \bibinfo {author} {\bibfnamefont {K.}~\bibnamefont {Hirose}}, \bibinfo
  {author} {\bibfnamefont {K.}~\bibnamefont {Haule}}, \bibinfo {author}
  {\bibfnamefont {K.}~\bibnamefont {Shimizu}}, \ and\ \bibinfo {author}
  {\bibfnamefont {Y.}~\bibnamefont {Ohishi}},\ }\href {\doibase
  10.1103/PhysRevLett.108.026403} {\bibfield  {journal} {\bibinfo  {journal}
  {Phys. Rev. Lett.}\ }\textbf {\bibinfo {volume} {108}},\ \bibinfo {pages}
  {026403} (\bibinfo {year} {2012})}\BibitemShut {NoStop}%
\bibitem [{\citenamefont {Gorelov}\ \emph {et~al.}(2010)\citenamefont
  {Gorelov}, \citenamefont {Karolak}, \citenamefont {Wehling}, \citenamefont
  {Lechermann}, \citenamefont {Lichtenstein},\ and\ \citenamefont
  {Pavarini}}]{r7}%
  \BibitemOpen
  \bibfield  {author} {\bibinfo {author} {\bibfnamefont {E.}~\bibnamefont
  {Gorelov}}, \bibinfo {author} {\bibfnamefont {M.}~\bibnamefont {Karolak}},
  \bibinfo {author} {\bibfnamefont {T.~O.}\ \bibnamefont {Wehling}}, \bibinfo
  {author} {\bibfnamefont {F.}~\bibnamefont {Lechermann}}, \bibinfo {author}
  {\bibfnamefont {A.~I.}\ \bibnamefont {Lichtenstein}}, \ and\ \bibinfo
  {author} {\bibfnamefont {E.}~\bibnamefont {Pavarini}},\ }\href {\doibase
  10.1103/PhysRevLett.104.226401} {\bibfield  {journal} {\bibinfo  {journal}
  {Phys. Rev. Lett.}\ }\textbf {\bibinfo {volume} {104}},\ \bibinfo {pages}
  {226401} (\bibinfo {year} {2010})}\BibitemShut {NoStop}%
\bibitem [{\citenamefont {Mravlje}\ \emph {et~al.}(2011)\citenamefont
  {Mravlje}, \citenamefont {Aichhorn}, \citenamefont {Miyake}, \citenamefont
  {Haule}, \citenamefont {Kotliar},\ and\ \citenamefont {Georges}}]{r8}%
  \BibitemOpen
  \bibfield  {author} {\bibinfo {author} {\bibfnamefont {J.}~\bibnamefont
  {Mravlje}}, \bibinfo {author} {\bibfnamefont {M.}~\bibnamefont {Aichhorn}},
  \bibinfo {author} {\bibfnamefont {T.}~\bibnamefont {Miyake}}, \bibinfo
  {author} {\bibfnamefont {K.}~\bibnamefont {Haule}}, \bibinfo {author}
  {\bibfnamefont {G.}~\bibnamefont {Kotliar}}, \ and\ \bibinfo {author}
  {\bibfnamefont {A.}~\bibnamefont {Georges}},\ }\href {\doibase
  10.1103/PhysRevLett.106.096401} {\bibfield  {journal} {\bibinfo  {journal}
  {Phys. Rev. Lett.}\ }\textbf {\bibinfo {volume} {106}},\ \bibinfo {pages}
  {096401} (\bibinfo {year} {2011})}\BibitemShut {NoStop}%
\bibitem [{\citenamefont {Zhang}\ \emph {et~al.}(2013)\citenamefont {Zhang},
  \citenamefont {Haule},\ and\ \citenamefont {Vanderbilt}}]{r9}%
  \BibitemOpen
  \bibfield  {author} {\bibinfo {author} {\bibfnamefont {H.}~\bibnamefont
  {Zhang}}, \bibinfo {author} {\bibfnamefont {K.}~\bibnamefont {Haule}}, \ and\
  \bibinfo {author} {\bibfnamefont {D.}~\bibnamefont {Vanderbilt}},\ }\href
  {\doibase 10.1103/PhysRevLett.111.246402} {\bibfield  {journal} {\bibinfo
  {journal} {Phys. Rev. Lett.}\ }\textbf {\bibinfo {volume} {111}},\ \bibinfo
  {pages} {246402} (\bibinfo {year} {2013})}\BibitemShut {NoStop}%
\bibitem [{\citenamefont {Haule}\ \emph {et~al.}(2014)\citenamefont {Haule},
  \citenamefont {Birol},\ and\ \citenamefont {Kotliar}}]{r10}%
  \BibitemOpen
  \bibfield  {author} {\bibinfo {author} {\bibfnamefont {K.}~\bibnamefont
  {Haule}}, \bibinfo {author} {\bibfnamefont {T.}~\bibnamefont {Birol}}, \ and\
  \bibinfo {author} {\bibfnamefont {G.}~\bibnamefont {Kotliar}},\ }\href
  {\doibase 10.1103/PhysRevB.90.075136} {\bibfield  {journal} {\bibinfo
  {journal} {Phys. Rev. B}\ }\textbf {\bibinfo {volume} {90}},\ \bibinfo
  {pages} {075136} (\bibinfo {year} {2014})}\BibitemShut {NoStop}%
\bibitem [{\citenamefont {Dang}\ \emph {et~al.}(2014)\citenamefont {Dang},
  \citenamefont {Ai}, \citenamefont {Millis},\ and\ \citenamefont
  {Marianetti}}]{r11}%
  \BibitemOpen
  \bibfield  {author} {\bibinfo {author} {\bibfnamefont {H.~T.}\ \bibnamefont
  {Dang}}, \bibinfo {author} {\bibfnamefont {X.}~\bibnamefont {Ai}}, \bibinfo
  {author} {\bibfnamefont {A.~J.}\ \bibnamefont {Millis}}, \ and\ \bibinfo
  {author} {\bibfnamefont {C.~A.}\ \bibnamefont {Marianetti}},\ }\href
  {\doibase 10.1103/PhysRevB.90.125114} {\bibfield  {journal} {\bibinfo
  {journal} {Phys. Rev. B}\ }\textbf {\bibinfo {volume} {90}},\ \bibinfo
  {pages} {125114} (\bibinfo {year} {2014})}\BibitemShut {NoStop}%
\bibitem [{\citenamefont {Dang}\ \emph {et~al.}(2015)\citenamefont {Dang},
  \citenamefont {Mravlje}, \citenamefont {Georges},\ and\ \citenamefont
  {Millis}}]{r12}%
  \BibitemOpen
  \bibfield  {author} {\bibinfo {author} {\bibfnamefont {H.~T.}\ \bibnamefont
  {Dang}}, \bibinfo {author} {\bibfnamefont {J.}~\bibnamefont {Mravlje}},
  \bibinfo {author} {\bibfnamefont {A.}~\bibnamefont {Georges}}, \ and\
  \bibinfo {author} {\bibfnamefont {A.~J.}\ \bibnamefont {Millis}},\ }\href
  {\doibase 10.1103/PhysRevB.91.195149} {\bibfield  {journal} {\bibinfo
  {journal} {Phys. Rev. B}\ }\textbf {\bibinfo {volume} {91}},\ \bibinfo
  {pages} {195149} (\bibinfo {year} {2015})}\BibitemShut {NoStop}%
\bibitem [{\citenamefont {Shorikov}\ \emph {et~al.}(2015)\citenamefont
  {Shorikov}, \citenamefont {Lukoyanov}, \citenamefont {Anisimov},\ and\
  \citenamefont {Savrasov}}]{bfodmft}%
  \BibitemOpen
  \bibfield  {author} {\bibinfo {author} {\bibfnamefont {A.~O.}\ \bibnamefont
  {Shorikov}}, \bibinfo {author} {\bibfnamefont {A.~V.}\ \bibnamefont
  {Lukoyanov}}, \bibinfo {author} {\bibfnamefont {V.~I.}\ \bibnamefont
  {Anisimov}}, \ and\ \bibinfo {author} {\bibfnamefont {S.~Y.}\ \bibnamefont
  {Savrasov}},\ }\href {\doibase 10.1103/PhysRevB.92.035125} {\bibfield
  {journal} {\bibinfo  {journal} {Phys. Rev. B}\ }\textbf {\bibinfo {volume}
  {92}},\ \bibinfo {pages} {035125} (\bibinfo {year} {2015})}\BibitemShut
  {NoStop}%
\bibitem [{\citenamefont {Kockelmann}\ \emph {et~al.}(2001)\citenamefont
  {Kockelmann}, \citenamefont {Sch{\"a}fer}, \citenamefont {Sosnowska},\ and\
  \citenamefont {Troyanchuk}}]{str}%
  \BibitemOpen
  \bibfield  {author} {\bibinfo {author} {\bibfnamefont {W.~A.}\ \bibnamefont
  {Kockelmann}}, \bibinfo {author} {\bibfnamefont {W.}~\bibnamefont
  {Sch{\"a}fer}}, \bibinfo {author} {\bibfnamefont {I.}~\bibnamefont
  {Sosnowska}}, \ and\ \bibinfo {author} {\bibfnamefont {I.}~\bibnamefont
  {Troyanchuk}},\ }in\ \href {\doibase
  10.4028/www.scientific.net/MSF.378-381.616} {\emph {\bibinfo {booktitle}
  {{European Powder Diffraction EPDIC 7}}}},\ \bibinfo {series} {{Mater. Sci.
  Forum}}, Vol.\ \bibinfo {volume} {378}\ (\bibinfo  {publisher} {Trans Tech
  Publications},\ \bibinfo {year} {2001})\ pp.\ \bibinfo {pages}
  {616--620}\BibitemShut {NoStop}%
\bibitem [{\citenamefont {Dreysse}(2000)}]{rspt1}%
  \BibitemOpen
  \bibinfo {editor} {\bibfnamefont {H.}~\bibnamefont {Dreysse}},\ ed.,\
  \href@noop {} {\emph {\bibinfo {title} {{Electronic Structure and Physical
  Properties of Solids: The Uses of the LMTO Method}}}}\ (\bibinfo  {publisher}
  {Springer Berlin Heidelberg},\ \bibinfo {year} {2000})\BibitemShut {NoStop}%
\bibitem [{\citenamefont {Wills}\ \emph {et~al.}(2010)\citenamefont {Wills},
  \citenamefont {Eriksson}, \citenamefont {Andersson}, \citenamefont {Delin},
  \citenamefont {Grechnyev},\ and\ \citenamefont {Alouani}}]{rspt2}%
  \BibitemOpen
  \bibfield  {author} {\bibinfo {author} {\bibfnamefont {J.~M.}\ \bibnamefont
  {Wills}}, \bibinfo {author} {\bibfnamefont {O.}~\bibnamefont {Eriksson}},
  \bibinfo {author} {\bibfnamefont {P.}~\bibnamefont {Andersson}}, \bibinfo
  {author} {\bibfnamefont {A.}~\bibnamefont {Delin}}, \bibinfo {author}
  {\bibfnamefont {O.}~\bibnamefont {Grechnyev}}, \ and\ \bibinfo {author}
  {\bibfnamefont {M.}~\bibnamefont {Alouani}},\ }\enquote {\bibinfo {title}
  {{Full-Potential Electronic Structure Method: Energy and Force Calculations
  with Density Functional and Dynamical Mean Field Theory}},}\ \ (\bibinfo
  {publisher} {Springer Berlin Heidelberg},\ \bibinfo {address} {Berlin,
  Heidelberg},\ \bibinfo {year} {2010})\BibitemShut {NoStop}%
\bibitem [{\citenamefont {von Barth}\ and\ \citenamefont {Hedin}(1972)}]{xc1}%
  \BibitemOpen
  \bibfield  {author} {\bibinfo {author} {\bibfnamefont {U.}~\bibnamefont {von
  Barth}}\ and\ \bibinfo {author} {\bibfnamefont {L.}~\bibnamefont {Hedin}},\
  }\href {http://stacks.iop.org/0022-3719/5/i=13/a=012} {\bibfield  {journal}
  {\bibinfo  {journal} {J. Phys. C: Solid State Phys.}\ }\textbf {\bibinfo
  {volume} {5}},\ \bibinfo {pages} {1629} (\bibinfo {year} {1972})}\BibitemShut
  {NoStop}%
\bibitem [{\citenamefont {Hedin}\ and\ \citenamefont {Lundqvist}(1971)}]{xc2}%
  \BibitemOpen
  \bibfield  {author} {\bibinfo {author} {\bibfnamefont {L.}~\bibnamefont
  {Hedin}}\ and\ \bibinfo {author} {\bibfnamefont {B.~I.}\ \bibnamefont
  {Lundqvist}},\ }\href {http://stacks.iop.org/0022-3719/4/i=14/a=022}
  {\bibfield  {journal} {\bibinfo  {journal} {J. Phys. C: Solid State Phys.}\
  }\textbf {\bibinfo {volume} {4}},\ \bibinfo {pages} {2064} (\bibinfo {year}
  {1971})}\BibitemShut {NoStop}%
\bibitem [{\citenamefont {Anisimov}\ \emph {et~al.}(1993)\citenamefont
  {Anisimov}, \citenamefont {Solovyev}, \citenamefont {Korotin}, \citenamefont
  {{Czy\ifmmode \dot{z}\else \.{z}\fi{}yk}},\ and\ \citenamefont
  {Sawatzky}}]{fll}%
  \BibitemOpen
  \bibfield  {author} {\bibinfo {author} {\bibfnamefont {V.~I.}\ \bibnamefont
  {Anisimov}}, \bibinfo {author} {\bibfnamefont {I.~V.}\ \bibnamefont
  {Solovyev}}, \bibinfo {author} {\bibfnamefont {M.~A.}\ \bibnamefont
  {Korotin}}, \bibinfo {author} {\bibfnamefont {M.~T.}\ \bibnamefont
  {{Czy\ifmmode \dot{z}\else \.{z}\fi{}yk}}}, \ and\ \bibinfo {author}
  {\bibfnamefont {G.~A.}\ \bibnamefont {Sawatzky}},\ }\href {\doibase
  10.1103/PhysRevB.48.16929} {\bibfield  {journal} {\bibinfo  {journal} {Phys.
  Rev. B}\ }\textbf {\bibinfo {volume} {48}},\ \bibinfo {pages} {16929}
  (\bibinfo {year} {1993})}\BibitemShut {NoStop}%
\bibitem [{\citenamefont {Caroli}\ \emph {et~al.}(1973)\citenamefont {Caroli},
  \citenamefont {Lederer-Rozenblatt}, \citenamefont {Roulet},\ and\
  \citenamefont {Saint-James}}]{caroli1973}%
  \BibitemOpen
  \bibfield  {author} {\bibinfo {author} {\bibfnamefont {C.}~\bibnamefont
  {Caroli}}, \bibinfo {author} {\bibfnamefont {D.}~\bibnamefont
  {Lederer-Rozenblatt}}, \bibinfo {author} {\bibfnamefont {B.}~\bibnamefont
  {Roulet}}, \ and\ \bibinfo {author} {\bibfnamefont {D.}~\bibnamefont
  {Saint-James}},\ }\href {\doibase 10.1103/PhysRevB.8.4552} {\bibfield
  {journal} {\bibinfo  {journal} {Phys. Rev. B}\ }\textbf {\bibinfo {volume}
  {8}},\ \bibinfo {pages} {4552} (\bibinfo {year} {1973})}\BibitemShut
  {NoStop}%
\bibitem [{\citenamefont {Feibelman}\ and\ \citenamefont
  {Eastman}(1974)}]{peter1974}%
  \BibitemOpen
  \bibfield  {author} {\bibinfo {author} {\bibfnamefont {P.~J.}\ \bibnamefont
  {Feibelman}}\ and\ \bibinfo {author} {\bibfnamefont {D.~E.}\ \bibnamefont
  {Eastman}},\ }\href {\doibase 10.1103/PhysRevB.10.4932} {\bibfield  {journal}
  {\bibinfo  {journal} {Phys. Rev. B}\ }\textbf {\bibinfo {volume} {10}},\
  \bibinfo {pages} {4932} (\bibinfo {year} {1974})}\BibitemShut {NoStop}%
\bibitem [{\citenamefont {Marksteiner}\ \emph {et~al.}(1986)\citenamefont
  {Marksteiner}, \citenamefont {Weinberger}, \citenamefont {Albers},
  \citenamefont {Boring},\ and\ \citenamefont {Schadler}}]{mark1986}%
  \BibitemOpen
  \bibfield  {author} {\bibinfo {author} {\bibfnamefont {P.}~\bibnamefont
  {Marksteiner}}, \bibinfo {author} {\bibfnamefont {P.}~\bibnamefont
  {Weinberger}}, \bibinfo {author} {\bibfnamefont {R.~C.}\ \bibnamefont
  {Albers}}, \bibinfo {author} {\bibfnamefont {A.~M.}\ \bibnamefont {Boring}},
  \ and\ \bibinfo {author} {\bibfnamefont {G.}~\bibnamefont {Schadler}},\
  }\href {\doibase 10.1103/PhysRevB.34.6730} {\bibfield  {journal} {\bibinfo
  {journal} {Phys. Rev. B}\ }\textbf {\bibinfo {volume} {34}},\ \bibinfo
  {pages} {6730} (\bibinfo {year} {1986})}\BibitemShut {NoStop}%
\bibitem [{\citenamefont {Panda}\ \emph {et~al.}(2016)\citenamefont {Panda},
  \citenamefont {Pal}, \citenamefont {Mandal}, \citenamefont {Gorgoi},
  \citenamefont {Das}, \citenamefont {Sarkar}, \citenamefont {Drube},
  \citenamefont {Sun}, \citenamefont {{Di Marco}}, \citenamefont {Lindblad},
  \citenamefont {Thunstr{\"o}m}, \citenamefont {Delin}, \citenamefont {Karis},
  \citenamefont {Kvashnin}, \citenamefont {van Schilfgaarde}, \citenamefont
  {Eriksson},\ and\ \citenamefont {Sarma}}]{panda2016}%
  \BibitemOpen
  \bibfield  {author} {\bibinfo {author} {\bibfnamefont {S.~K.}\ \bibnamefont
  {Panda}}, \bibinfo {author} {\bibfnamefont {B.}~\bibnamefont {Pal}}, \bibinfo
  {author} {\bibfnamefont {S.}~\bibnamefont {Mandal}}, \bibinfo {author}
  {\bibfnamefont {M.}~\bibnamefont {Gorgoi}}, \bibinfo {author} {\bibfnamefont
  {S.}~\bibnamefont {Das}}, \bibinfo {author} {\bibfnamefont {I.}~\bibnamefont
  {Sarkar}}, \bibinfo {author} {\bibfnamefont {W.}~\bibnamefont {Drube}},
  \bibinfo {author} {\bibfnamefont {W.}~\bibnamefont {Sun}}, \bibinfo {author}
  {\bibfnamefont {I.}~\bibnamefont {{Di Marco}}}, \bibinfo {author}
  {\bibfnamefont {A.}~\bibnamefont {Lindblad}}, \bibinfo {author}
  {\bibfnamefont {P.}~\bibnamefont {Thunstr{\"o}m}}, \bibinfo {author}
  {\bibfnamefont {A.}~\bibnamefont {Delin}}, \bibinfo {author} {\bibfnamefont
  {O.}~\bibnamefont {Karis}}, \bibinfo {author} {\bibfnamefont {Y.~O.}\
  \bibnamefont {Kvashnin}}, \bibinfo {author} {\bibfnamefont {M.}~\bibnamefont
  {van Schilfgaarde}}, \bibinfo {author} {\bibfnamefont {O.}~\bibnamefont
  {Eriksson}}, \ and\ \bibinfo {author} {\bibfnamefont {D.~D.}\ \bibnamefont
  {Sarma}},\ }\href {\doibase 10.1103/PhysRevB.93.235138} {\bibfield  {journal}
  {\bibinfo  {journal} {Phys. Rev. B}\ }\textbf {\bibinfo {volume} {93}},\
  \bibinfo {pages} {235138} (\bibinfo {year} {2016})}\BibitemShut {NoStop}%
\end{thebibliography}
%

\end{document}